\documentclass[11pt,a4paper]{article}
\usepackage{jheppub}
\usepackage{amsfonts}
\usepackage{amsmath}
\usepackage{amsbsy}
\usepackage{latexsym}
\usepackage{pdflscape}
\usepackage{bm}
\numberwithin{equation}{section}
\allowdisplaybreaks

\DeclareMathOperator{\Tr}{Tr}
\DeclareMathOperator{\tr}{tr}
\DeclareMathOperator{\Str}{Str}
\DeclareMathOperator{\Det}{Det}

\renewcommand{\thefootnote}{\fnsymbol{footnote}}
\newcommand{\nn}{\nonumber \\}

\newcommand{\bra}{\langle}
\newcommand{\ket}{\rangle}
\def\vev#1{\langle #1 \rangle}
\def\({\Bigl(}
\def\){\Bigr)}

\def\cO{{\mathcal O}}
\def\l{\ell}

\preprint{DESY\ 13-097, TIT/HEP-628, KEK-TH-1636, YITP-13-57}

\title{ABJM Wilson Loops in Arbitrary Representations}

\author[a,b]{Yasuyuki Hatsuda,}
\author[c,d]{Masazumi Honda,}
\author[d,e]{Sanefumi Moriyama}
\author[f]{and Kazumi Okuyama}

\affiliation[a]{DESY Theory Group, DESY Hamburg, \\
Notkestrasse 85, D-22603 Hamburg, Germany}

\affiliation[b]{Department of Physics, \\
Tokyo Institute of Technology, Tokyo 152-8551, Japan}
\affiliation[c]{High Energy Accelerator Research Organization (KEK), \\
Tsukuba, Ibaraki 305-0801, Japan}
\affiliation[d]{ Yukawa Institute for Theoretical Physics, Kyoto University,\\
Kitashirakawa Oiwakecho, Sakyo-ku, Kyoto 606-8502, Japan} 
\affiliation[e]{Kobayashi Maskawa Institute
and Graduate School of Mathematics, \\
Nagoya University, Nagoya 464-8602, Japan}
\affiliation[f]{Department of Physics, \\
Shinshu University, Matsumoto 390-8621, Japan}

\emailAdd{yasuyuki.hatsuda@desy.de} 
\emailAdd{mhonda@post.kek.jp}
\emailAdd{moriyama@math.nagoya-u.ac.jp} 
\emailAdd{kazumi@azusa.shinshu-u.ac.jp}

\abstract{
We study vacuum expectation values (VEVs) of circular half BPS Wilson
loops in arbitrary representations in ABJM theory.
We find that those in hook representations are reduced to elementary
integrations thanks to the Fermi gas formalism, which are accessible
from the numerical studies similar to the partition function in the
previous studies.
For non-hook representations, we show that the VEVs in the grand
canonical formalism can be exactly expressed as determinants of those
in the hook representations.
Using these facts, we can study the instanton effects of the VEVs in
various representations.
Our results are consistent with the worldsheet instanton effects
studied from the topological string and a prescription to include the
membrane instanton effects by shifting the chemical potential, which
has been successful for the partition function.
}

\begin{document}

\maketitle

\renewcommand{\thefootnote}{\arabic{footnote}}
\setcounter{footnote}{0}
\setcounter{section}{0}

\section{Introduction}
Recently, there has been much progress in understanding membranes in
M-theory.
It was proposed in \cite{ABJM} that the low energy effective theory on
the $N$ multiple M2-branes on the geometry
${\mathbb C}^4 /{\mathbb Z}_k$ is described by the 3-dimensional
${\cal N}=6$ supersymmetric generalization of the Chern-Simons matter
theory with gauge group $U(N)_k\times U(N)_{-k}$ commonly referred as
ABJM theory.
Furthermore it has been shown by using the localization technique
\cite{P} that a class of supersymmetric observables in the ABJM theory
on $S^3$ are described by so-called ABJM matrix model
\cite{KWY,J,HHL,Drukker:2012sr}.

The partition function $Z(N)$ is the first fundamental quantity to be
studied.
After the rather standard matrix model analysis in
\cite{DMP1,DMP2,FHM}, there appeared a seminal paper, which rewrites
the ABJM partition function into the partition function of an ideal
Fermi gas system \cite{MP2} (see also \cite{HKPT,O,Kapustin:2010xq}).
One of the advantages in this Fermi gas formalism is that instead of
the stringy 't Hooft expansion, we can access to the M-theory region
directly by taking large $N$ limit with $k$ fixed.
As is usual in the statistical system, instead of the partition
function, it is convenient to define the grand partition function
\begin{align}
\Xi(z)=\sum_{N=0}^\infty z^NZ(N),
\label{Xi}
\end{align}
by introducing the fugacity $z=e^\mu$ with the chemical potential
$\mu$.
Subsequently in \cite{KEK,HMO1,PY,HMO2,CM,HMO3,HMMO}, the partition
function of the ABJM theory was studied extensively from this grand
partition function of the Fermi gas system.
Finally, it turned out that the grand potential $J(\mu)=\log\Xi(z)$
can be separated into the perturbative, worldsheet instanton
\cite{Cagnazzo:2009zh}, membrane instanton \cite{Becker:1995kb,DMP2}
and bound state part.
The worldsheet instanton part is determined directly from the
topological string result \cite{HMO2}.
The membrane instanton part is also related to the refined topological
string \cite{HMMO}.
As found in \cite{HMO3}, the contributions from all of the bound
states can be incorporated to the worldsheet instanton effects by
shifting the chemical potential $\mu$ to an ``effective'' chemical
potential $\mu_{\rm eff}$, which is described by the sum of $\mu$ and
a part of the pure membrane instanton effects.

Here we proceed to study the second fundamental quantity, namely, the
vacuum expectation value (VEV) of the circular half BPS Wilson loop\footnote{
Below we often refer to this circular half BPS Wilson loop simply as ``the half BPS Wilson loop".
As seen later in \eqref{eq:W-VEV}, after applying the localization method,
the Wilson loop operator becomes a character of a certain group representation in mathematical terminology,
to which we mostly continue to refer as the Wilson loop by a slight abuse of terminology.
}
firstly introduced in \cite{DT,MP1}.
The half BPS Wilson loops have nice counterparts in the open
topological string, as was pointed out in \cite{MP1,DMP1}.
This is one of our motivation that we focus on them here.
The half BPS Wilson loops are classified by representations
$\mathbf{R}$ of the supergroup $U(N|N)$, which includes the gauge
group $U(N)\times U(N)$ as the bosonic subgroup.
By using the localization method \cite{KWY,J,HHL,Drukker:2012sr}, the
unnormalized VEV of the Wilson loop $W_{\mathbf{R}}$ in the
representation $\mathbf{R}$ is written as
\begin{eqnarray}
&&\bra W_{\mathbf{R}} \ket_N
=\frac{1}{(N!)^2}\int\prod_i\frac{d\mu_i}{2\pi}\frac{d\nu_i}{2\pi}
\frac{\prod_{i<j}(2\sinh\frac{\mu_i-\mu_j}{2})^2
(2\sinh\frac{\nu_i-\nu_j}{2})^2}
{\prod_{i,j}(2\cosh\frac{\mu_i-\nu_j}{2})^2}
e^{-\frac{1}{2g_s}\sum_i (\mu_i^2 -\nu_i^2)} \Str_{\mathbf{R}}U ,\nonumber \\
&& U=\begin{pmatrix}U_\mu&0\\0& -U_\nu\end{pmatrix},\quad
U_\mu={\rm diag}(e^{\mu_i}),\quad U_\nu={\rm diag}(e^{\nu_i}) ,
\label{eq:W-VEV}
\end{eqnarray}
where $g_s =\frac{2\pi i}{k}$ is the coupling constant, and
$\Str_{\mathbf{R}}$ is the $U(N|N)$ character in the representation
$\mathbf{R}$.
A prescription to obtain $\Str_{\mathbf{R}}$ is summarized as follows.
First, a representation of the supergroup $U(N|N)$ is characterized
by the super Young diagram, which has the same form as the usual Young
diagram of the bosonic group $U(\infty)$ (for example, see
\cite{Bars:1982ps}).
Then, the supertrace $\Str_{\mathbf{R}} U$ of the supergroup $U(N|N)$
is found if we formally replace the power sum $\tr U^n$ in
$\tr_{\mathbf{R}} U$ of  $U(\infty)$ by $\Str U^n$.
Note that $U$ appearing in $\Str_{\mathbf{R}} U$ is a $2N \times 2N$
matrix defined by \eqref{eq:W-VEV}.

The computation of the VEVs using the Fermi gas formalism was
initiated in \cite{KMSS}, where the inserted observables are
restricted to the operators with winding number $n$, $\Str U^n$.
Very recently, it was proposed in \cite{GKM} that it is possible to
study the perturbative part and the worldsheet instanton part using
the topological strings.
This subject keeps on attracting various studies.\footnote{
See, for example, \cite{BGLP} for perturbative studies of the Wilson
loop VEVs, \cite{Farquet:2013cwa,K} for the holographic studies,
\cite{K,Griguolo:2012iq,Cardinali:2012ru} for generalizations of contour and
\cite{S} for more general Chern-Simons matter theory.}

In this paper, we present a Fermi gas formalism for the VEVs in
arbitrary representations, suitable for numerical study, and study
these non-perturbative effects.
As in the partition function, besides the worldsheet instanton
contribution, 
we also find the contribution coming from the membrane instanton,
which is difficult to be known from the topological string theory.
In the following of this introduction, we would like to explain our
results in more details.
Just as in the partition function, it is useful to consider 
the VEV in the grand canonical ensemble defined by
\begin{align}
\bra W_{\mathbf{R}} \ket^{\rm GC}
=\frac{1}{\Xi(z)}\sum_{N=0}^\infty z^N\bra W_{\mathbf{R}} \ket_N.
\label{OGC}
\end{align}
Note that once we know $\bra W_{\mathbf{R}}\ket^{\rm GC}$, the VEVs
in the canonical ensemble is easily recovered.

\begin{figure}[tb]
\begin{center}
\begin{tabular}{cc}
\resizebox{60mm}{!}{\includegraphics{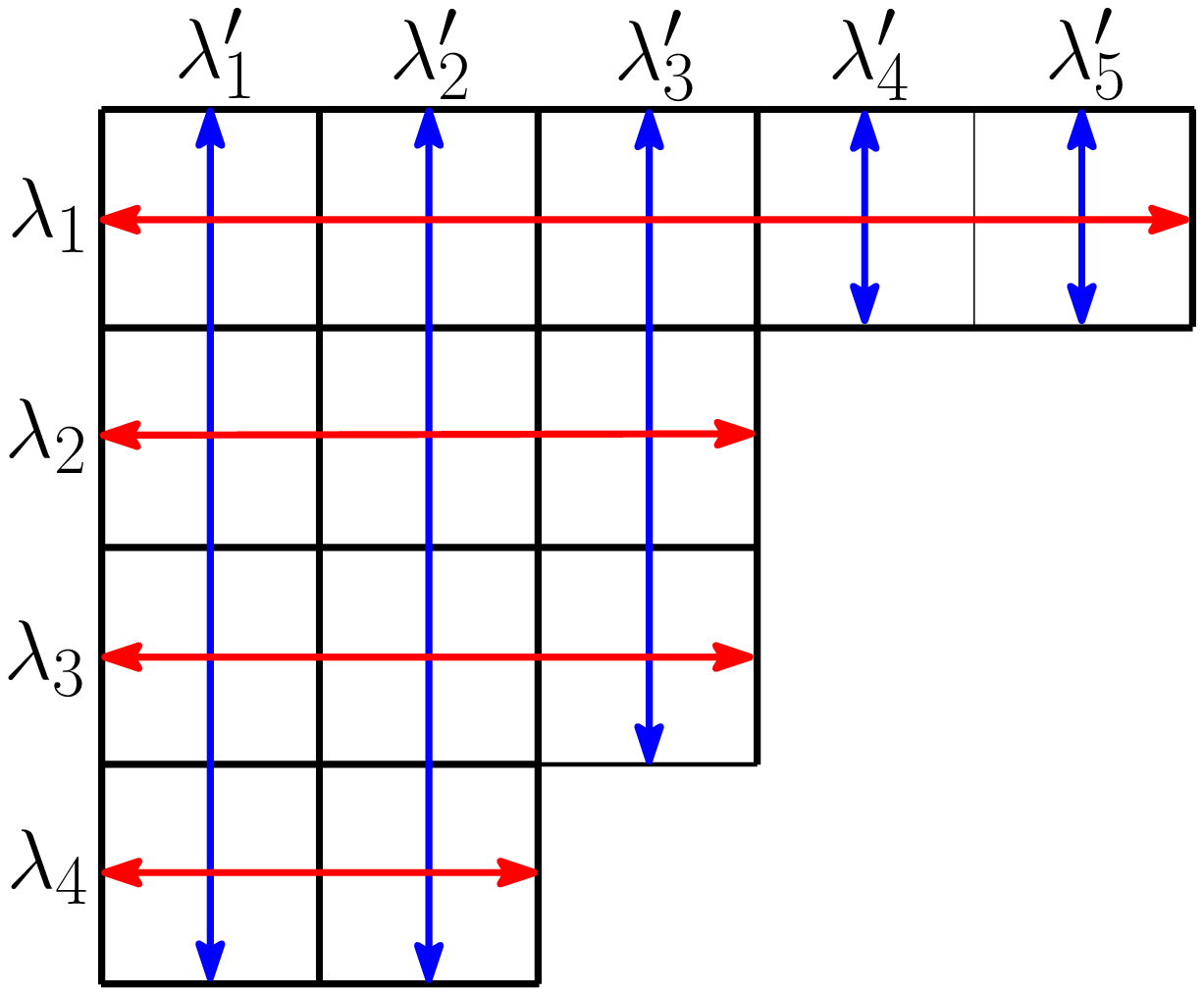}}
&
\resizebox{60mm}{!}{\includegraphics{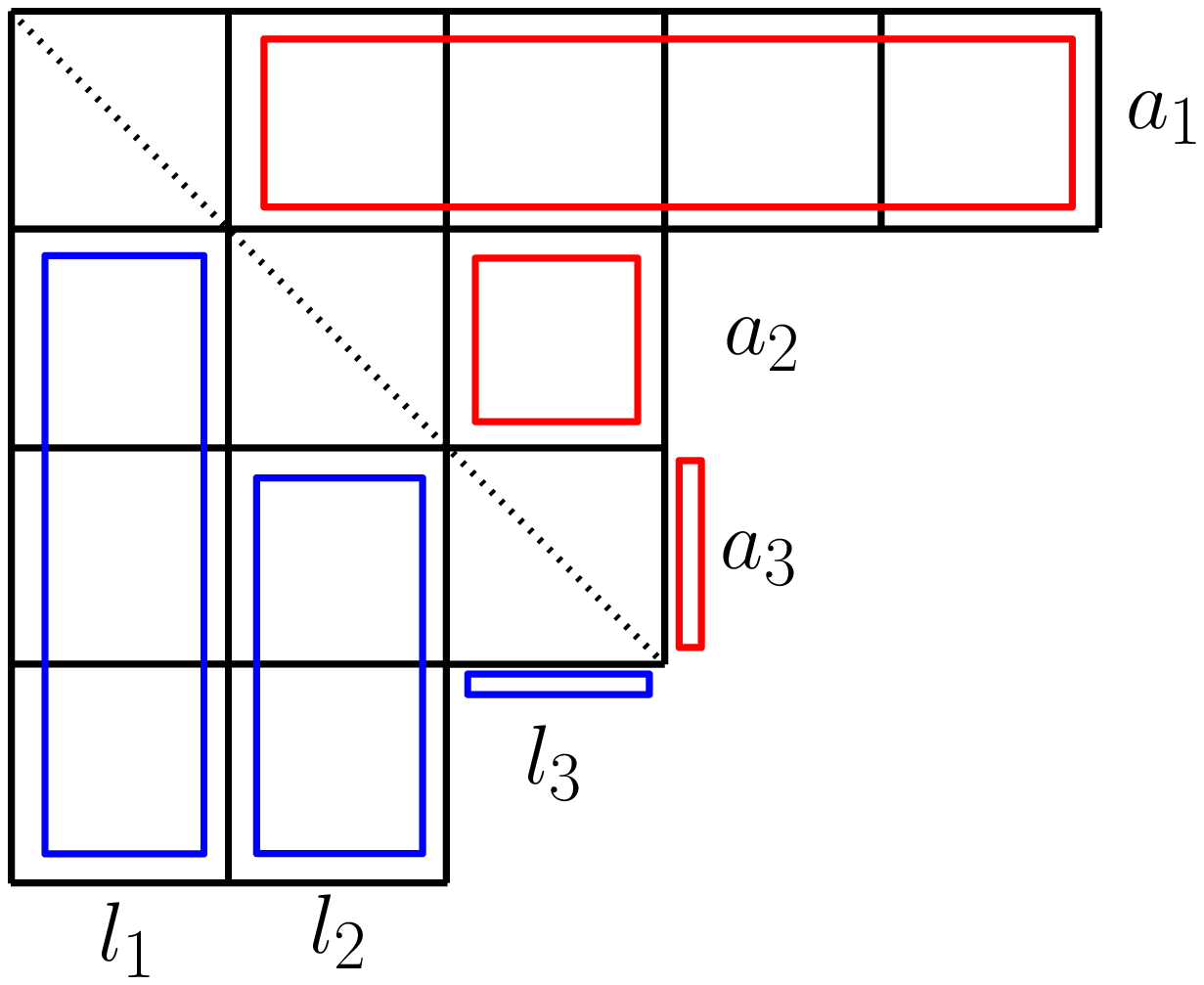}}
\\
(a) Partition notation & (b) Frobenius notation 
\end{tabular}
\end{center}
\caption{
(a) The partition notation $[\lambda_1\lambda_2\lambda_3\cdots]$ with
its transpose $[\lambda'_1\lambda'_2\lambda'_3\cdots]$ and (b) the
Frobenius notation $(a_1a_2\cdots a_r|l_1l_2\cdots l_r)$ for the same
Young diagram.
Here $r=\max\{s|\lambda_s-s\ge 0\}=\max\{s|\lambda'_s-s\ge 0\}$ is the
number of diagonal boxes, and $a_p$, $l_q$ denote the horizontal and
vertical distances from each diagonal box, respectively, given by
$a_p=\lambda_p-p,\ l_q=\lambda'_q-q$.
In the above case, the Young diagram is
$[\lambda_1\lambda_2\lambda_3\lambda_4]=[5,3,3,2]$ in the partition
notation with its transpose
$[\lambda'_1\lambda'_2\lambda'_3\lambda'_4\lambda'_5]=[4,4,3,1,1]$,
while it is $(a_1a_2 a_3 | l_1l_2 l_3 )=(4,1,0|3,2,0)$ in the
Frobenius notation.}
\label{fig:notation}
\end{figure}
First, we find a formula for the VEV of the Wilson loop in the hook
representation\footnote{
Throughout this paper, we use the Frobenius notation to express
representations of $U(N|N)$ illustrated in Figure \ref{fig:notation}.
} $\mathbf{R}=(a|l)$ in terms of a certain convolution of integrations
\begin{align}
\langle W_{(a|l)}\rangle^{\rm GC}=\langle a|\frac{z}{1+z\rho_1}|l\rangle .
\label{hook}
\end{align}
Here $\rho_1$ is the density operator of the Fermi-gas system defined
later in \eqref{eq:rho1}, and the states $\langle a|$ and $|l\rangle$
in the coordinate basis are given by \eqref{ndef}.
The expression \eqref{hook} is accessible from the numerical studies
with very high precision.

Second, we extend our analysis to general representation
${\mathbf R}=(a_1a_2\cdots a_r|l_1l_2\cdots l_r)$.
In operator level, the Wilson loop is simply given by
the determinant of those in the hook representations, known as the
Giambelli formula,\footnote{
We often write $W_{\mathbf{R}}(e^\mu,e^\nu)$ instead of
${\rm Str}_{\mathbf{R}} U$ to represent the Wilson loop insertion
apparently.}
\begin{align}
W_{(a_1a_2\cdots a_r|l_1l_2\cdots l_r)}(e^{\mu},e^{\nu})
=\det{}_{p,q}W_{(a_p|l_q)}(e^{\mu},e^{\nu}) .
\end{align}
In this paper, we find that the VEVs in the grand canonical ensemble
exactly satisfy the same type of the formula,
\begin{align}
\langle W_{(a_1a_2\cdots a_r|l_1l_2\cdots l_r)}\rangle^{\rm GC}
=\langle \det{}_{p,q}  W_{(a_p|l_q)} \rangle^{\rm GC}
=\det{}_{p,q}\Big(\langle W_{(a_p|l_q)}\rangle^{\rm GC}\Big).
\label{nonhook}
\end{align}
Hence the VEVs of the half BPS Wilson loops in general representations
can be computed only from those in hook representations.
We emphasize that this result is very unexpected and non-trivial.
In the mathematical context, a normalized linear functional
$\langle{\mathcal O}\rangle$ of symmetric functions ${\mathcal O}$
satisfying the above property is called Giambelli compatible
(see e.g.~\cite{BOS}).
Let us further call a linear functional being factorizable if it
satisfies the property
$\langle{\mathcal O}_1{\mathcal O}_2\rangle
=\langle{\mathcal O}_1\rangle\langle{\mathcal O}_2\rangle$.
Note that the factorizability implies the Giambelli compability.
In this terminology, we show that the grand canonical  VEV of the half
BPS Wilson loop is Giambelli compatible but not factorizable.
We also find that its perturbative part is factorizable (see
\eqref{eq:factorization}).
Note that the factorization of the grand canonical VEV also implies
that of the canonical VEV in the large $N$ limit, which is natural
from the physical viewpoint.
The factorization property, however, is generically broken by the
instanton contributions.
Nevertheless, the Giambelli compatibility is still preserved after the
instanton effects are taken into account.

Finally, using our results (\ref{hook}) and (\ref{nonhook}),
we also study the structure of the instanton corrections to the VEVs
in various representations by the numerical studies.
The VEVs, in general, receive the following corrections,
\begin{align}
\vev{W_\mathbf{R}}^{\rm GC}
=W_\mathbf{R}^{\rm GC(pert)}(\mu,k)
(1+W_\mathbf{R}^{\rm GC(WS)}(\mu,k)+W_\mathbf{R}^{\rm GC(others)}(\mu,k)),
\end{align}
where $W_\mathbf{R}^{\rm GC(pert)}(\mu,k)$ is the perturbative part,
$W_\mathbf{R}^{\rm GC(WS)}(\mu,k)$ is the worldsheet instanton
correction, and $W_\mathbf{R}^{\rm GC(others)}(\mu,k)$ consists of the
pure membrane instanton correction and the contribution from the bound
states.
We have found that our numerical results match with the topological
string prediction of the perturbative part and the worldsheet
instanton part with the chemical potential shifted from $\mu$ to
$\mu_{\rm eff}$ to incorporate the contribution from the membrane
instantons and the bound states:
\begin{align}
\vev{W_\mathbf{R}}^{\rm GC}
=W_\mathbf{R}^{\rm GC(pert)}(\mu_{\rm eff},k)
(1+W_\mathbf{R}^{\rm GC(WS)}(\mu_{\rm eff},k)),
\label{eq:Wnp}
\end{align}
exactly the same as in the partition function.
Here the ``effective'' chemical potential $\mu_{\rm eff}$ was
introduced in \cite{HMO3} in order to explain the bound state
contribution in the grand potential,
\begin{align}
\mu_{\rm eff}=\mu+\frac{\pi^2 k}{2} \sum_{\ell=1}^\infty a_\ell(k) e^{-2\ell \mu},
\label{eq:mu-mu_eff}
\end{align}
where $a_\ell(k)$ are the functions appearing in the membrane
instanton correction in the grand potential.
The forms of $a_\ell(k)$ are exactly computed by the refined
topological string on local $\mathbb{P}^1 \times \mathbb{P}^1$
\cite{HMMO}.
We should stress that the perturbative part and the worldsheet
instanton part in \eqref{eq:Wnp} are computed from the open
topological string on local $\mathbb{P}^1 \times \mathbb{P}^1$ as we
will see in section 5.
Thus our result states that once we determine the topological string
free energy on this background, we can exactly find the VEVs of the
half BPS Wilson loops in general representations in the ABJM theory.

The organization of this paper is as follows.
In the next section we present a general framework to study the VEVs
of the BPS Wilson loops and apply it to the half BPS case in the hook
representation in section 3.
Since it is difficult to apply this formalism directly to the half BPS
Wilson loops in the non-hook representation, we shall present an
alternative method in section 4, which works only for the half BPS
Wilson loop.
After reviewing the results from the topological strings in section 5,
we summarize our numerical study in section 6.
Finally we conclude in section 7.

\section{BPS Wilson loops in general representations}
Here we present methods to study the VEV of the
Wilson loop in the ABJM theory using the Fermi gas formalism.
We shall first present a framework to study general $1/6$ BPS Wilson
loop constructed in \cite{1_6BPS}, which includes the half BPS Wilson
loop as a special case.

\subsection{Partition function}
For this purpose let us first review the derivation of the Fermi gas
formalism for the partition function \cite{MP2} carefully because our
Wilson loop insertion is based heavily on it.
The starting point is the partition function of the ABJM matrix model
\cite{KWY,J,HHL,Drukker:2012sr}:
\begin{align}
Z(N)
=\frac{1}{(N!)^2}\int\prod_i\frac{d\mu_i}{2\pi}\frac{d\nu_i}{2\pi}
\frac{\prod_{i<j}(2\sinh\frac{\mu_i-\mu_j}{2})^2
(2\sinh\frac{\nu_i-\nu_j}{2})^2}
{\prod_{i,j}(2\cosh\frac{\mu_i-\nu_j}{2})^2}
e^{-\frac{1}{2g_s}(\sum_i\mu_i^2-\sum_j\nu_j^2)},
\label{PF}
\end{align}
By using the Cauchy identity and performing a Fourier transformation, 
the partition function \eqref{PF} is rewritten into
\begin{align}
&Z(N)=\frac{1}{N!}\int\prod_i\frac{dx_i}{\hbar}\frac{dy_i}{\hbar}
\sum_{\sigma\in S_N}(-1)^\sigma\nonumber\\
&\quad\times\int\prod_i\frac{dp_i}{2\pi}\frac{dq_i}{2\pi}
\prod_i \biggl[ \frac{e^{-\frac{i}{\hbar}p_i(x_i-y_i)}}{2\cosh\frac{p_i}{2}}
\frac{e^{-\frac{i}{\hbar}q_i(x_i-y_{\sigma^{-1}(i)})}}{2\cosh\frac{q_i}{2}}\biggr]
e^{\frac{i}{2\hbar}(\sum_ix_i^2-\sum_jy_j^2)},
\label{firstZ}
\end{align}
where we rescale the integration variables as
$\mu_i=\frac{x_i}{k},\ \nu_j=\frac{y_j}{k}$ and $\hbar=2\pi k$.
After performing the Gaussian integral over $x$ and $y$ by completing
the square in the exponent
\begin{align}
&\frac{i}{2\hbar}x_i^2-\frac{i}{\hbar}x_i(p_i+q_i)
-\frac{i}{2\hbar}y_i^2+\frac{i}{\hbar}y_i(p_i+q_{\sigma(i)})
\nonumber\\
&\quad=\frac{i}{2\hbar}(x_i-p_i-q_i)^2
-\frac{i}{2\hbar}(y_i-p_i-q_{\sigma(i)})^2
-\frac{i}{2\hbar}(p_i+q_i)^2+\frac{i}{2\hbar}(p_i+q_{\sigma(i)})^2,
\end{align}
and noting the cancellation of the $p^2$ and $q^2$ terms, the
partition function becomes
\begin{align}
Z(N)=\frac{1}{N!}\int\prod_i\frac{dp_idq_i}{2\pi\hbar}
\sum_{\sigma\in S_N}(-1)^\sigma\prod_i
\frac{e^{ip_i(q_{\sigma(i)}-q_i)/\hbar}}
{2\cosh\frac{p_i}{2}\cdot 2\cosh\frac{q_i}{2}}.
\label{Zrho}
\end{align}
If we further integrate over $p$ in \eqref{Zrho}, then we find
\begin{align}
Z(N)=\frac{1}{N!}\int\prod_i\frac{dq_i}{\hbar}\sum_{\sigma\in S_N}(-1)^\sigma
\prod_i
\frac{1}{\sqrt{2\cosh\frac{q_{\sigma(i)}}{2}}}
\frac{1}{2\cosh\frac{q_{\sigma(i)}-q_i}{2k}}
\frac{1}{\sqrt{2\cosh\frac{q_i}{2}}}.
\end{align}
Since the partition function $Z(N)$ has the form of an ideal Fermi gas
system as
\begin{align}
Z(N)=\frac{1}{N!}\sum_{\sigma\in S_N}(-1)^\sigma
\int\prod_i\frac{dq_i}{\hbar}\prod_i \rho_1 (q_i,q_{\sigma(i)}),
\label{density}
\end{align}
with
\begin{align}
\rho_1(q_i,q_j)=
\frac{1}{\sqrt{2\cosh\frac{q_j}{2}}}
\frac{1}{2\cosh\frac{q_j-q_i}{2k}}
\frac{1}{\sqrt{2\cosh\frac{q_i}{2}}},
\end{align}
it is easier to consider the grand canonical partition function
\eqref{Xi} by introducing the fugacity $z=e^\mu$.
One can show that the grand partition function is expressed as a
Fredholm determinant,
\begin{align}
\Xi(z)=\Det(1+z\rho_1),
\label{Fredholm}
\end{align}
where the determinant $\Det$ is taken over the whole Hilbert space of
the Fermi gas system.
In the operator formalism, the density matrix $\rho_1$ is given by
\begin{align}
\rho_1=\sqrt{Q}P\sqrt{Q},
\quad {\rm with}\quad 
P=\frac{1}{2\cosh\frac{p}{2}},\ 
Q=\frac{1}{2\cosh\frac{q}{2}}, 
\label{eq:rho1}
\end{align}
where $q$ and $p$ satisfies the canonical commutation relation
$[q,p]=i\hbar$ with $\hbar=2\pi k$.
We adopt this notation in what follows.

\subsection{Operator insertion}
General $1/6$ BPS Wilson loops in the ABJM theory are generated by the
following type of operator \cite{KWY}:
\begin{equation}
\prod_i f(e^{\mu_i})g(e^{\nu_i}),
\end{equation}
where $f(x)$ and $g(x)$ are functions of $x$.
In this section we translate the insertion of this operator into the
one of a certain quantum mechanical operator expressed by $(q,p)$.

As a warm up, let us first consider the operator insertion
\begin{align}
e^{n\mu_M}=e^{\frac{2\pi nx_M}{\hbar}},
\end{align}
into the partition function (\ref{firstZ}).
After completing the square in integrating over $x_M$ and combining
with the contribution from integrating $y_M$ as in the computation of
the partition function, we find an extra contribution into the
exponent:
\begin{align}
\frac{i}{\hbar}p_M(-2\pi i n)
+\frac{2\pi n}{\hbar}(q_M+\pi in).
\end{align}
Performing the integration over $p_M$, the unnormalized VEV is finally
given by
\begin{align}
\langle e^{n\mu_M}\rangle_N
&=\frac{1}{N!}\int\prod_i\frac{dq_i}{\hbar}\sum_{\sigma\in S_N} (-1)^\sigma
    \prod_{i\ne M} \rho_1 (q_{\sigma (i)},q_i ) \nonumber \\
&~~~~~~~~ \times \frac{1}{\sqrt{2\cosh\frac{q_{\sigma(M)}}{2}}}
\frac{e^{\frac{2\pi n}{\hbar}(q_M+\pi in)}}
{2\cosh\frac{q_{\sigma(M)}-q_M-2\pi in}{2k}}
\frac{1}{\sqrt{2\cosh\frac{q_M}{2}}} .
\end{align}
In the language of quantum mechanical operators, the second line can
be interpreted as the matrix element
\begin{align}
\langle q_{\sigma(M)}|\sqrt{Q}Pe^{\frac{n(q+p)}{k}}\sqrt{Q}|q_M\rangle
=\langle q_{\sigma(M)}|\sqrt{Q}e^{\frac{np}{k}}
Pe^{\frac{n(q+i\pi n)}{k}}\sqrt{Q}|q_M\rangle .
\end{align}
Therefore we conclude that the insertion of the operator $e^{n\mu_M}$
amounts to the insertion of the operator $W^n$ to the right of $P$, where $W$ is defined by
\begin{align}
W=e^{\frac{q+p}{k}}.
\end{align}

Similarly, we find that the insertion of the operator $e^{n\nu_M}$
amounts to insertion of the same operator $W^n$ to the left of $P$.
This can be seen by repeating the square completion in the exponent
with an extra factor
\begin{align}
\frac{i}{\hbar}p_M(-2\pi in)
+\frac{2\pi n}{\hbar}(q_{\sigma(M)}-\pi in),
\end{align}
and computing of the matrix element
\begin{align}
\langle q_{\sigma(i)}|\sqrt{Q}e^{\frac{n(q+p)}{k}}P\sqrt{Q}|q_i\rangle
=\langle q_{\sigma(i)}|\sqrt{Q}e^{\frac{n(q-\pi in)}{k}}
e^{\frac{np}{k}}P\sqrt{Q}|q_i\rangle.
\end{align}
Note that this interpretation is factor-wise.
Namely, not only other additive terms in the insertion do not affect
this interpretation, but this interpretation is valid even if this
operator is multiplied by other operators.
We can also see that the simultaneous insertion at the same position $M$,
namely, $e^{m\mu_M+n\nu_M}$ also works well.

Therefore we can summarize the computation rule as follows.
For the case of the partition function, we finally end up with the
summation over the conjugacy classes and the study of
\begin{align}
\Tr\rho_1^m=\Tr\sqrt{Q}PQPQPQP\cdots\sqrt{Q}.
\end{align}
For the case of Wilson loop, we insert $W$ into various slots
between $Q$ and $P$ in this trace.
The insertion pattern depends on the representation, but since we are
considering the gauge invariant operator, we have to take a trace,
namely, sum over all the insertion slots.
Hence our formula can be summarized as
\begin{align}
\Xi(z)\Big\langle\prod_if(e^{\mu_i})g(e^{\nu_i})\Big\rangle^{\rm GC}
=\Det\Big(1+z\sqrt{Q}g(W)Pf(W)\sqrt{Q}\Big),
\label{grandWins}
\end{align}
where $\bra{\cal O}\ket^{\rm GC}$ denotes the expectation value
of the operator ${\cal O}$ in the grand canonical ensemble
\eqref{OGC}.
Once the grand canonical VEV is understood, one can easily return to
the canonical VEV via
\begin{equation}
\bra {\cal O} \ket_N = \frac{1}{2\pi i}
\oint \frac{dz}{z^{N+1}}\ \Xi (z) \bra {\cal O} \ket^{\rm GC} .
\label{eq:GC-C}
\end{equation}

Alternatively, we can show the relation \eqref{grandWins} using
the operator formalism as follows.
The expectation value of $\prod_if(e^{\mu_i})g(e^{\nu_i})$ at fixed
$N$ is given by
\begin{align}
&\Big\langle\prod_if(e^{\mu_i})g(e^{\nu_i})\Big\rangle_N
=\frac{1}{N!}\sum_{\sigma\in S_N}(-1)^{\sigma}
\int\prod_i\frac{d\mu_i}{2\pi}\frac{d\nu_i}{2\pi}
\nn
&\quad\times\prod_if(e^{\mu_i})g(e^{\nu_i})e^{\frac{ik}{4\pi}(\mu^2_i-\nu^2_i)}
\prod_{i}\frac{1}{2\cosh\frac{\nu_{\sigma(i)} -\mu_{i}}{2}}
\frac{1}{2\cosh\frac{\mu_i -\nu_i}{2}}.
\end{align}
By rescaling $\mu_i=\frac{x_i}{k},\nu_i=\frac{y_i}{k}$, this is
rewritten as
\begin{align}
\Big\langle\prod_if(e^{\mu_i})g(e^{\nu_i})\Big\rangle_N
=\frac{1}{N!}\sum_{\sigma\in S_N}(-1)^{\sigma}\int \prod_i \frac{dy_i}{\hbar}
\prod_{i}\rho(y_i,y_{\sigma(i)}),
\end{align}
where $\rho$ denotes the density matrix in the presence of 
operator insertion
\begin{align}
\rho (y_i,y_j)&=\int\frac{dx}{\hbar}
\frac{e^{\frac{i}{2\hbar}(x^2 -y_j^2)}f(e^{\frac{x}{k}})g(e^{\frac{y_j}{k}})}
{2\cosh\frac{y_j -x}{2k}\cdot2\cosh\frac{x -y_i}{2k}}\nn
&=\int \frac{dx}{\hbar}
\bra y_j|e^{-\frac{iq^2}{2\hbar}} g(e^{\frac{q}{k}})
\frac{1}{2\cosh\frac{p}{2}}|x\ket
\bra x|f(e^{\frac{q}{k}}) e^{\frac{iq^2}{2\hbar}}
\frac{1}{2\cosh\frac{p}{2}}|y_i\ket\nn
&=\bra y_j|e^{-\frac{iq^2}{2\hbar}} g(e^{\frac{q}{k}})
\frac{1}{2\cosh\frac{p}{2}}f(e^{\frac{q}{k}}) e^{\frac{iq^2}{2\hbar}}
\frac{1}{2\cosh\frac{p}{2}}|y_i\ket.
\label{fgrho}
\end{align}
This can be written as an operator equation
\begin{align}
\rho
&= e^{-\frac{iq^2}{2\hbar}} g(e^{\frac{q}{k}})
\frac{1}{2\cosh\frac{p}{2}}f(e^{\frac{q}{k}}) e^{\frac{iq^2}{2\hbar}}
\frac{1}{2\cosh\frac{p}{2}} \nn
&=e^{-\frac{iq^2}{2\hbar}} e^{-\frac{ip^2}{2\hbar}}g(e^{\frac{q+p}{k}})\frac{1}{2\cosh\frac{p}{2}}
f(e^{\frac{q+p}{k}})\frac{1}{2\cosh\frac{q}{2}}
e^{\frac{ip^2}{2\hbar}} e^{\frac{iq^2}{2\hbar}},
\label{densityfg}
\end{align}
where we have used
\begin{align}
 e^{\frac{iq^2}{2\hbar}}F(p)e^{-\frac{iq^2}{2\hbar}}
=F(p-q),\quad
e^{\frac{ip^2}{2\hbar}}G(q)e^{-\frac{ip^2}{2\hbar}}
=G(q+p).
\end{align}
Therefore, up to a similarity transformation the density matrix in
\eqref{densityfg} becomes
\begin{align}
\rho=\sqrt{Q}g(W)Pf(W)\sqrt{Q},
\label{hrhoAn}
\end{align}
which reproduces \eqref{grandWins}.

\section{Half BPS Wilson loops I: hook representations}\label{sec:hook}
In the previous section, we have presented a general framework to
study the VEVs of the general 1/6 BPS Wilson loop 
in the Fermi gas formalism.
Especially we have reduced the problem into computing the trace with
alternating operators $Q$ and $P$ and various $W$-insertions.
This quantity, however, is still difficult to compute, at least,
numerically with high precision.
Here we would like to see what kind of simplification will occur if
we restrict ourselves to the half BPS Wilson loops.

\subsection{Representations of the superalgebra}
The half BPS Wilson loop is classified by the representation of
$U(N|N)$ \cite{DT,MP1}.
In this subsection we review representations of the supergroup
$U(N|N)$.
For this purpose, it is convenient to consider representations of
$U(\infty)$.
A simple prescription to derive the character of $U(N|N)$ is to
formally replace $\tr U^n$ in the character $\tr_{\mathbf{R}}U$ of
$U(\infty)$ by $\Str U^n$:
\begin{equation}
{\rm Str}_{\mathbf{R}} U = 
\left. {\rm tr}_{\mathbf{R}} U \right|_{{\rm tr}U^n \rightarrow {\rm Str}U^n} .
\end{equation}
Note that the character $\tr_{\mathbf{R}}U$ is given by the Schur
function associated with the Young diagram $\mathbf{R}$.
The supertrace ${\rm Str}_{\mathbf{R}} U$ can be expressed by a
combination of characters of two bosonic subgroups $U(N)$ of
$U(N|N)$.
For example, in the case of the 2nd anti-symmetric representation
$(0|1)$, the superalgebraic generalization turns out to be
\begin{align}
{\rm Str}_{(0|1)} U
&=\frac{1}{2}(\Str{ U})^2 -\frac{1}{2}\Str{ U}^2\nn
&=\tr_{(0|1)} U_\mu +\tr_{(0|0)} U_\mu \tr_{(0|0)} U_\nu +\tr_{(1|0)} U_\nu.
\end{align}
where $U_{\mu}$ and $U_{\nu}$ are the bosonic parts of $U$ (see
\eqref{eq:W-VEV}).
Below, we often denote the supertrace ${\rm Str}_{\mathbf{R}} U$ by
$W_{\mathbf{R}}(e^\mu,e^\nu)$, and use the abbreviation
$W_{\mathbf{R}}=W_{\mathbf{R}}(e^\mu,e^\nu)$ as long as there is no
risk of confusion.

\subsection{Beyond winding Wilson loops}
The Wilson loop with the winding number $n$ 
\begin{align}
\Str U^n=\sum_ie^{n\mu_i}-(-1)^n\sum_ie^{n\nu_i},
\end{align}
was studied extensively in \cite{KMSS}.
By revisiting this in our formalism, we will obtain a hint to study the
more general representations as in the following.

In our formalism, applying the rule in \eqref{grandWins} with the
choice,
\begin{align}
f(W)=1+tW^n,\quad
g(W)=\frac{1}{1+t(-W)^n}=\frac{1}{f(-W)},
\end{align}
and picking up the linear term in $t$, the grand canonical VEV of
$\Str U^n$ is given by
\begin{align}
\bra\Str U^n\ket^{\rm GC}
=\Tr\left[R(z)\sqrt{Q}\left(PW^n-(-W)^nP\right)\sqrt{Q}\right],
\label{winding}
\end{align} 
with $R(z)$ defined by
\begin{align}
R(z)=\frac{z}{1+z\rho_1}.
\end{align}
One can easily see that the operator appearing on the right-hand-side
is expanded as
\begin{align}
PW^n-(-W)^nP= \sum_{l=0}^{n-1}(-1)^lW^l(WP+PW)W^{n-1-l}.
\label{sumhook}
\end{align}
Note that the operator appearing in the right-hand-side of
\eqref{sumhook} has the factorized form
\begin{align}
\langle q_2|W^{n}(WP+PW)W^{m}|q_1\rangle
=\langle q_2|\frac{1}{\sqrt{Q}}|n\rangle
\langle m|\frac{1}{\sqrt{Q}}|q_1\rangle,
\label{factor}
\end{align}
where the coordinate $q$ representations of $|n\rangle$ and
$\langle m|$ are defined by
\begin{align}
\langle q|n\rangle
=\frac{e^{(n+\frac{1}{2})\frac{q}{k}-\frac{\pi i}{k}n(n+1)}}
{\sqrt{2\cosh\frac{q}{2}}},\quad
\langle m|q\rangle=\langle q|m\rangle^*
=\frac{e^{(m+\frac{1}{2})\frac{q}{k}+\frac{\pi i}{k}m(m+1)}}
{\sqrt{2\cosh\frac{q}{2}}}.
\label{ndef}
\end{align}
As a formal operator relation, \eqref{factor} is also written as
\begin{align}
\sqrt{Q}W^{n}(WP+PW)W^{m}\sqrt{Q}=|n\rangle\langle m|.
\end{align}
Thus we finally obtain the grand canonical VEV of the winding Wilson
loop \eqref{winding} as
\begin{align}
\bra\Str U^n\ket^{\rm GC}
=\sum_{l=0}^{n-1}(-1)^l\bra n-1-l|R(z)|l\ket.
\end{align}
Comparing with the relation between the winding Wilson loop $\Str U^n$
and the Wilson loop $W_{(a|l)}$ in the hook representation
\begin{align}
\Str U^n=\sum_{l=0}^{n-1}(-1)^lW_{(n-1-l|l)},
\end{align}
it is tantalizing to expect the relation
\begin{align}
\bra W_{(a|l)}\ket^{\rm GC}=\bra a|R(z)|l\ket,
\label{hookloop}
\end{align}
which is true as we will see in the next subsection.

More generally, the computation of the VEVs of the half BPS operators
reduces to picking up a certain function $f(W)$ and computing the
Fredholm determinant of the corresponding density matrix $\rho_f$
\begin{align}
\Xi(z)
\Big\langle\prod_i\frac{f(e^{\mu_i})}{f(-e^{\nu_i})}\Big\rangle^{\rm GC}
=\Det(1+z\rho_f) \quad
{\rm with}\quad
\rho_f=\sqrt{Q}\frac{1}{f(-W)}Pf(W)\sqrt{Q}.
\label{halfBPSloop}
\end{align}
Rewriting the density matrix in the above expression as
\begin{align}
\rho_f-\rho_1
&=\sqrt{Q}\frac{1}{f(-W)}\Big(Pf(W)-f(-W)P\Big)\sqrt{Q}\nn
&=\sqrt{Q}\frac{1}{f(-W)}
\sum_{n=0}^\infty\frac{f^{(n)}(0)}{n!}(PW^n-(-W)^nP)\sqrt{Q},
\end{align}
and recalling \eqref{sumhook}, one can see that the grand canonical
VEV of the half BPS Wilson loops can always be written as a sum of the
factorized functions.

\subsection{Single-hook representations}
For the half BPS Wilson loop in a single-hook representation $(a|l)$, 
the generating function is given by \cite{hook}
\begin{align}
1+(s+t)\sum_{a,l=0}^\infty s^at^lW_{(a|l)}
={\rm Sdet}\left(\frac{1+t U}{1-s U}\right)=
\prod_{j=1}^N\frac{(1+te^{\mu_j})(1+se^{\nu_j})}{(1-se^{\mu_j})(1-te^{\nu_j})}.
\label{sdet}
\end{align}
When plugging
\begin{align}
f(W)=\frac{1+tW}{1-sW},
\end{align}
into our formula \eqref{halfBPSloop}, we find that the corresponding
density matrix factorizes as
\begin{align}
\rho_f&=\rho_1+(s+t)\sum_{a,l=0}^\infty s^at^l|l\ket\bra a|.
\end{align}
Therefore, the grand canonical VEV of \eqref{sdet} becomes
\begin{align}
\Big\langle 
&1+(s+t)\sum_{a,l=0}^\infty s^at^lW_{(a|l)}\Big\rangle^{\rm GC}
=\frac{\Det(1+z\rho_f)}{\Det(1+z\rho_1)}\nn
&=\Det\left(1+(s+t)\sum_{a,l=0}^\infty s^at^l
R(z)|l\ket\bra a|\right)=1+(s+t)\sum_{a,l=0}^\infty s^at^l
\bra a|R(z)|l\ket.
\end{align}
Finally, the grand canonical VEV of $W_{(a|l)}$ is found to be
\eqref{hookloop} which is accessible from the numerical studies
similar to the partition function in the previous studies
\cite{HMO1,PY,HMO2,HMO3}.

\section{Half BPS Wilson loops II: general representations}\label{sec:nonhook}
In the previous sections, we have presented a method to compute the
supersymmetric Wilson loops and shown that especially for the half BPS
Wilson loop in the hook representation, there is a factorization,
which at least simplifies the numerical study.
The above analysis for the hook representation is, however, difficult
to be extended to a general non-hook case.
Here we shall present a completely different analysis which is
effective for studying the non-hook representations from the hook
representations but only suitable for the half BPS Wilson loop.

\subsection{Non-hook representations}
After understanding the VEV in the hook representation in the previous
section, we can go beyond the hook representation step by step.
Namely, we can substitute various functions for $f(W)$ and subtract
the known hook part.
For example, if we plug $f(W)=e^{tW}$, which corresponds to the
generating function of $({\rm Str}U)^n$, and compare
$\mathcal{O}(t^4)$ terms, then we find
\begin{align}
\langle W_{(10|10)}\rangle^{\rm GC}
=\det
\begin{pmatrix}\langle W_{(1|1)}\rangle^{\rm GC}
&\langle W_{(1|0)}\rangle^{\rm GC}\\
\langle W_{(0|1)}\rangle^{\rm GC}
&\langle W_{(0|0)}\rangle^{\rm GC} .
\end{pmatrix}
=\det\begin{pmatrix}\langle 1|R(z)|1\rangle
&\langle 1|R(z)|0\rangle\\
\langle 0|R(z)|1\rangle
&\langle 0|R(z)|0\rangle
\end{pmatrix}
\end{align}

More generally, it is easy to imagine the expression in
\eqref{nonhook}.
By changing the function for $f(W)$, we will encounter various
relations supporting this conjecture.
However, it is difficult to prove it directly using this formulation.

\subsection{A proof}
Here we give a proof of \eqref{nonhook}: 
\[
\langle W_{(a_1a_2\cdots a_r|l_1l_2\cdots l_r)}\rangle^{\rm GC}
=\det{}_{p,q}\Big(\langle W_{(a_p|l_q)}\rangle^{\rm GC}\Big) ,
\]
with a completely different
method.\footnote{We are grateful to Sho Matsumoto for his
collaborative contribution in sharing his idea of proof and the
references with us in this subsection.}
The Giambelli formula states that
\[
W_{(a_1a_2\cdots a_r|l_1l_2\cdots l_r)}(e^{\mu},e^{\nu})
=\det{}_{p,q}W_{(a_p|l_q)}(e^{\mu},e^{\nu}).
\]
Therefore, we would like to study
\begin{align}
\langle W_{(a_1a_2\cdots a_r|l_1l_2\cdots l_r)} (e^{\mu},e^{\nu}) \rangle
=\langle\det{}_{p,q}W_{(a_p|l_q)}(e^{\mu},e^{\nu})\rangle.
\end{align} 
Instead of computing it directly, here let us consider
\begin{align}
W(N)=\langle\det{}_{p,q}(\delta_{p,q}+tW_{(a_p|l_q)}(e^{\mu},e^{\nu}))\rangle,
\end{align}
and picking up the coefficient of the highest $t^r$ term.
The reason we want to consider $W(N)$ is because this is a
generalization of the Cauchy determinant
\begin{align}
\frac{\det{}_{i,j}[(x_i+y_j)^{-1}
+t\sum_{p=1}^rx_i^{a_p}y_j^{l_p}]}
{\det{}_{i,j}[(x_i+y_j)^{-1}]}
=\det{}_{p,q}[\delta_{p,q}+tW_{(a_p|l_q)}(x,y)].
\label{Cauchy}
\end{align}
The proof of this formula for $r=1$ is simply reduced to a more
general formula in \cite{MVdJ}.\footnote{The formula of \cite{MVdJ} for the $r=1$ case is written as
\begin{align}
W_{(a|l)}(x,y)=\sum_{i,j=1}^N y_i^lM^{-1}_{ij}x_j^a
\label{hookschur}
\end{align}
where $M^{-1}$ is the inverse of Cauchy matrix $M_{ij}=1/(x_i+y_j)$.
One can show that the generating
function of \eqref{hookschur} reproduces \eqref{sdet}. 
}
The proof for $r>1$ is reduced to the case of $r=1$ by the formula
\begin{align}
\det{}_{I,J=1,\cdots,N}
\Big(\delta_{I,J}+\sum_{k=1}^n(V)_{Ik}(U^{\rm T})_{kJ}\Big)
=\det{}_{i,j=1,\cdots,n}
\Big(\delta_{i,j}+\sum_{K=1}^N(U^{\rm T})_{iK}(V)_{Kj}\Big),
\label{det=det}
\end{align}
which is true since $\tr(VU^{\rm T})^m=\tr(U^{\rm T}V)^m$ for any
positive integer $m$.
To simplify our computation in the following, let us define
\begin{align}
[d\mu_i]=\frac{d\mu_i}{2\pi}e^{-\frac{1}{2g_s}\mu_i^2},\quad
[d\nu_i]=\frac{d\nu_i}{2\pi}e^{\frac{1}{2g_s}\nu_i^2}.
\label{measure}
\end{align}
Then the quantity we want to compute becomes
\begin{align}
W(N)
&=\frac{1}{N!}
\int\prod_i[d\mu_i][d\nu_i]
\det{}_{p,q}(\delta_{p,q}+tW_{(a_p,l_q)}(e^{\mu},e^{\nu})) \nonumber\\
& \times \sum_{\sigma\in S_N}(-1)^\sigma 
 \prod_i\frac{1}{2\cosh\frac{\nu_{\sigma(i)}-\mu_i}{2}}
\frac{1}{2\cosh\frac{\mu_i-\nu_i}{2}}.
\end{align}
Using the formula \eqref{Cauchy}, we can rewrite this as
\begin{align}
W(N)
&=\frac{1}{N!}\sum_{\sigma\in S_N}(-1)^\sigma
\int\prod_i[d\nu_i]\prod_i\bm\rho(\nu_i,\nu_{\sigma(i)} ),
\end{align}
where
\begin{equation}
\bm\rho(\nu_i,\nu_j)
=\int[d\mu]
\Big(\frac{1}{2\cosh\frac{\nu_j-\mu}{2}}
+t\sum_{p=1}^re^{(l_p+1/2)\nu_j}e^{(a_p+1/2)\mu}\Big)
\frac{1}{2\cosh\frac{\mu-\nu_i}{2}} .
\end{equation}
Since the VEV can be interpreted as the partition function of the
ideal Fermi gas system just as the partition function \eqref{density},
it is natural to introduce the generating function as
\begin{align}
\Omega(z)=\sum_{N=0}^\infty z^NW(N) ={\rm \mathbf{Det}} (1+z\bm{\rho}),
\end{align}
where ${\rm \mathbf{Det}}$ is defined through the trace over the indices $\nu$ with the measure in \eqref{measure}.
Therefore, if we define
\begin{align}
&\bm Q(\mu, \nu)
=\frac{1}{2\cosh\frac{\mu -\nu}{2}},\quad
\bm P(\nu , \mu )
=\frac{1}{2\cosh\frac{\nu-\mu}{2}},\quad
\bm{\rho_1}=\sqrt{\bm Q}\bm P\sqrt{\bm Q}, \nonumber\\
&\left( \langle\bm a|\frac{1}{\sqrt{\bm Q}} \right) (\mu ) =e^{(a+1/2)\mu},\quad
\left( \frac{1}{\sqrt{\bm Q}}|\bm l\rangle \right) (\nu ) =e^{(l+1/2)\nu},\quad
\end{align}
then we find
\begin{align}
\Omega(z)&=
{\rm \mathbf{Det}}\Big(1
+z\Big(\bm P
+t\sum_{p=1}^r\frac{1}{\sqrt{\bm Q}}|\bm{l_p}\rangle
\langle\bm{a_p}|\frac{1}{\sqrt{\bm Q}}\Big)
\bm Q\Big)
\nonumber\\
&={\rm \mathbf{Det}}(1+z\bm{\rho_1})
\det{}_{p,q}\big(\delta_{p,q}
+zt\langle\bm{a_p}|
(1+z\bm{\rho_1})^{-1}|\bm{l_q}\rangle\big) .
\end{align}
where the multiplication among variables in the boldface character are
understood as matrix multiplication with indices $\mu,\nu$ and
measures in \eqref{measure}.
Note that the square root $\sqrt{\bm Q}$ should be regarded as a
formal notation.
We can express the integrations without it.
The reason we introduce it is because of the relation to the previous
quantities as we shall see below.
Here, in the last equation we have used the formula
\begin{align}
\det{}_{i,j=1,\cdots,D}\Big(\delta_{i,j}+\sum_{p=1}^r(l_p)_i(a_p)_j\Big)
=\det{}_{p,q=1,\cdots,r}\Big(\delta_{p,q}+\sum_{i=1}^D(a_p)_i(l_q)_i\Big),
\end{align}
which is the same as \eqref{det=det} if we change the variables by
$(V)_{Ik}=(l_k)_I$, $(U^{\rm T})_{kJ}=(a_k)_J$.
Now if we pick up the $t^r$ term, then we find
\begin{align}
\Xi(z)\Big\langle\det{}_{p,q}W_{(a_p|l_q)}(e^{\mu},e^{\nu})\Big\rangle^{\rm GC}
={\rm \mathbf{Det}}(1+z\bm{\rho_1})
\det{}_{p,q}
\langle\bm{a_p}|z(1+z\bm{\rho_1})^{-1}|\bm{l_q}\rangle.
\label{Giambelli}
\end{align}
This holds for both the hook and the non-hook cases.

Now using this result \eqref{Giambelli} we can reduce the proof of
\eqref{nonhook} to the result of \eqref{hook} given in the previous
section or we can prove \eqref{hook} independently.
Let us first consider to reduce to the previous result.
If we pick up the constant term by taking the limit $t\to 0$, we find
\begin{align}
\sum_{N=0}^\infty z^N\langle 1\rangle_N
={\rm \mathbf{Det}}(1+z\bm{\rho_1}).
\end{align}
Comparing with the expression for the partition function
\eqref{Fredholm}, we find
\begin{align}
{\rm \mathbf{Det}}(1+z\bm\rho_1)=\Det(1+z\rho_1).
\label{pf}
\end{align}
Also, if we apply the above results to the single-hook case, we find
\begin{align}
\Xi(z)\langle W_{(a_p|l_q)}(e^{\mu},e^{\nu})\rangle^{\rm GC}
={\rm \mathbf{Det}}(1+z\bm{\rho_1})
z\langle\bm{a_p}|(1+z\bm{\rho_1})^{-1}|\bm{l_q}\rangle.
\end{align}
Again comparing with the expression for the hook representation we
have, we find
\begin{align}
\langle\bm{a_p}|(1+z\bm{\rho_1})^{-1}|\bm{l_q}\rangle
=\langle a_p|(1+z\rho_1)^{-1}|l_q\rangle.
\label{hk}
\end{align}
Plugging \eqref{pf} and \eqref{hk} back to \eqref{Giambelli}, we have
shown that
\begin{align}
\Big\langle
W_{(a_1a_2\cdots a_r|l_1l_1\cdots l_r)}\Big\rangle^{\rm GC}
=\det{}_{p,q}\Big(\langle a_p|R(z)|l_q\rangle\Big).
\label{hookdet-formula}
\end{align}

Instead of our comparison with the known results, the argument here
also suggests that if we restrict ourselves to the half BPS Wilson
loop, we can have an alternative derivation for the hook case if we
evaluate carefully ${\rm \mathbf{Det}}(1+z\bm{\rho_1})$ and
$\langle \bm{a_p}|(1+z\bm{\rho_1})^{-1}|\bm{l_q}\rangle$.
The computation of ${\rm \mathbf{Det}}(1+z\bm{\rho_1})$ is exactly what we did
around \eqref{Zrho}.
Also, the computation of
$\langle \bm{a}|(1+z\bm{\rho_1})^{-1}| \bm{l}\rangle$
becomes
\begin{align}
\int\frac{dx}{\hbar}e^{ix^2/(2\hbar)}
\cdots\int\frac{dy}{\hbar}e^{-iy^2/(2\hbar)}
e^{2\pi(a+\frac{1}{2})x/\hbar}\frac{e^{-iq_x(x-y')/\hbar}}{2\cosh\frac{q_x}{2}}
\cdots\frac{e^{-iq_y(x'-y)/\hbar}}{2\cosh\frac{q_y}{2}}e^{2\pi(l+\frac{1}{2})y/\hbar},
\end{align}
In completing the square for $x$ and $y$ we find
\begin{align}
\frac{i}{2\hbar}x^2-\frac{iq_x}{\hbar}x+\frac{2\pi(a+1/2)}{\hbar}x
&=\frac{i}{2\hbar}(x-q_x-2\pi i(a+1/2))^2
-\frac{i}{2\hbar}(q_x+2\pi i(a+1/2))^2,
\nonumber\\
-\frac{i}{2\hbar}y^2+\frac{iq_y}{\hbar}y+\frac{2\pi(l+1/2)}{\hbar}y
&=-\frac{i}{2\hbar}(y-q_y+2\pi i(l+1/2))^2
+\frac{i}{2\hbar}(q_y-2\pi i(l+1/2))^2.
\end{align}
Note that $q^2$ terms cancel with the square completion from the
neighboring terms.
Hence, we are left with
\begin{align}
\frac{1}{2k}((a+1/2)q_x+2\pi ia(a+1)+(l+1/2)q_y-2\pi il(l+1)).
\end{align}
This is nothing but the exponent we found in \eqref{factor} with
\eqref{ndef}.
We note in passing that the above computation can be done also in the
operator formalism.

\subsection{Fermionic representation}
Our general expression \eqref{hookdet-formula} of the Wilson loop VEV
suggests that there is an underlying fermionic structure.
This is expected from the
fermionic nature of D-branes in topological string theory
\cite{Aganagic:2003qj}. 
Introducing the fermions
\begin{align}
\psi(x)=\sum_{n\in \mathbb{Z}}\psi_{n+\frac{1}{2}}x^{-n-1},\quad
\psi^*(x)=\sum_{n\in \mathbb{Z}}\psi^*_{n+\frac{1}{2}}x^{-n-1},
\end{align} 
with the standard anti-commutation relation
\begin{align}
\{\psi_r,\psi_s^*\}=\delta_{r+s,0},
\end{align}
such that the vacuum is annihilated by the positive modes as
\begin{equation}
\psi_r |0\rangle = \psi_r^* |0\rangle =0\quad {\rm for}\quad r>0,
\end{equation}
we define the state $|V\ket$ as
\begin{align}
|V\ket=\exp\left[\sum_{a,l=0}^\infty\bra W_{(a|l)}\ket^{\rm GC}
\psi_{-a-\frac{1}{2}}\psi^*_{-l-\frac{1}{2}}\right]|0\ket
=\exp\left[\sum_{a,l=0}^\infty\bra a|R(z)|l\ket
\psi_{-a-\frac{1}{2}}\psi^*_{-l-\frac{1}{2}}\right]|0\ket.
\end{align}
In terms of this state $|V\ket$, the grand canonical VEV of the Wilson
loop $W_{(a_1a_2\cdots a_r|l_1l_2\cdots l_r)}$ is compactly written as
\begin{align}
\Big\langle
W_{(a_1a_2\cdots a_r|l_1l_1\cdots l_r)}\Big\rangle^{\rm GC}
= \bra 0|\prod_{i=1}^r\psi^*_{a_i+\frac{1}{2}}\psi_{l_i+\frac{1}{2}}|V\ket.
\end{align} 
This is reminiscent of the expression of topological vertex in
\cite{Aganagic:2003qj}.
Indeed, the perturbative part of a single-hook Wilson loop is
determined by  the topological vertex of $\mathbb{C}^3$
\begin{align}
\bra W_{(a|l)}\ket^{\rm GC(pert)}
&=\frac{q^{\frac{1}{4} a(a+1)-\frac{1}{4} l(l+1)}}{[a+l+1][a]![l]!}
i^{a+l+1}e^{\frac{2(a+l+1)\mu}{k}},
\label{hookpert}
\end{align}
with $[n]=q^{\frac{n}{2}}-q^{-\frac{n}{2}}$ and $q=e^{\frac{4\pi i}{k}}$.
Using the $q$-binomial formula, one can show that
the alternating sum of \eqref{hookpert} 
reproduces the perturbative part of
winding Wilson loop \cite{KMSS}
\begin{align}
 \sum_{a+l=n-1}(-1)^l
\frac{q^{\frac{1}{4} a(a+1)-\frac{1}{4} l(l+1)}}{[a+l+1][a]![l]!}i^{a+l+1}e^{\frac{2(a+l+1)\mu}{k}}
=\frac{i^n}{[n]}e^{\frac{2n\mu}{k}}
=\frac{i^{n-1}}{2\sin\frac{2\pi}{k}}e^{\frac{2n\mu}{k}}.
\end{align}

\section{Relation to open topological strings}
In this section, we see a relation between the VEVs of the half BPS
Wilson loops and the open topological string amplitudes. 
As is well-known, the ABJM matrix model is related to the $L(2,1)$
lens space matrix model by analytic continuation \cite{MP1,DMP1} (see
also \cite{Analytic,AHS}).
This lens space matrix model is also related to the topological string
on local $\mathbb{P}^1 \times \mathbb{P}^1$ through the large $N$
duality \cite{Aganagic:2002wv}.
In fact, the perturbative and the worldsheet instanton parts in the
ABJM partition function can be captured by the result of the closed
topological string on local $\mathbb{P}^1 \times \mathbb{P}^1$.
Similarly, the VEVs of the half BPS Wilson loops are described by the
open topological string.
Here we are interested in the VEVs in the grand canonical ensemble,
which corresponds to the so-called large radius frame on the
topological string side.
The open topological string in this frame was recently studied in
detail in \cite{GKM}.

We note that the membrane instanton corrections are difficult to be
known from the topological string because these corrections correspond
to the non-perturbative effects in the topological string.
We will explore the membrane instanton corrections in the next section
with the help of the numerical analysis.

First we briefly summarize the result of \cite{GKM}.
The open topological string amplitudes take the following general 
form \cite{Ooguri:1999bv,Labastida:2000yw,Marino:2001re},
\begin{align}
F^{\rm open}(t,V)
=\sum_{\beta \in H_2(X)}\sum_{g=0}^\infty\sum_{h=1}^\infty
\sum_{\boldsymbol{\ell}}\sum_{m=1}^\infty
\frac{1}{h!}n_{g,\beta,\boldsymbol{\ell}}
\frac{1}{m}\( 2\sinh \frac{m g_{\rm top}}{2}\)^{2g-2}\nn
\times \prod_{j=1}^h
\(\frac{2}{\ell_j}\sinh\frac{m\ell_j g_{\rm top}}{2}
\tr V^{m\ell_j}\) e^{-m \beta \cdot t} ,
\label{eq:open_top}
\end{align}
where $t$ is the K\"ahler moduli of the local Calabi-Yau $X$, and $V$
is the open string moduli.
For the ABJM theory, we are interested in local
$\mathbb{P}_1\times\mathbb{P}_1$.
The string coupling in the topological string is related to the
Chern-Simons level,
\begin{align}
g_{\rm top}=\frac{4\pi i}{k}.
\end{align}
There are two K\"ahler moduli, which are identified as the chemical
potential $\mu$ dual to the original rank $N$, 
\begin{align}
t_1=t_2=T=\frac{4\mu}{k}-\pi i,\quad Q\equiv e^{-T}=-e^{-\frac{4\mu}{k}}.
\end{align}
Similarly, the open string moduli $V$ is also identified with the dual
variable for the Wilson loop insertion $U$.
Then, we can relate the perturbative part and worldsheet instanton
part of the grand canonical VEVs in the ABJM theory to the above open
topological string amplitudes.
The concrete relation is given explicitly by \cite{GKM}, 
\begin{align}
\label{eq:open-GC}
e^{F^{\rm open}(t,\widehat{V})}
&=\biggl\langle\exp
\biggl[\sum_{j=1}^\infty\frac{1}{j}\Str U^j\tr V^j\biggr]
\biggr\rangle^{\rm GC(pert+WS)}\\
&=\sum_{n_1,n_2,\dots }c_{n_1,n_2,\dots}
\vev{(\Str U)^{n_1}(\Str U^2)^{n_2}\cdots}^{\rm GC(pert+WS)}
(\tr V)^{n_1}(\tr V^2)^{n_2}\cdots,\notag
\end{align}
with $c_{n_1,n_2,\dots}=1/(\prod_{j}j^{n_j}n_j!)$.
Note that to write down the relation we have to plug a new parameter
\begin{align}
\widehat{V}=Q^{-1/2}V=ie^{\frac{2\mu}{k}}V
\label{eq:hV-V},
\end{align}
into \eqref{eq:open_top}.

\subsection{Perturbative part}
Let us consider the perturbative part.
We neglect all the exponentially suppressed terms in
\eqref{eq:open_top}.
We observe that the leading order contribution $\beta=(0,0)$ comes
only from
\begin{align}
n_{0,(0,0),(1)}=1.
\end{align}
Thus we obtain
\begin{align}
F^{\rm open}_{\rm pert}(V)=\frac{1}{i}
\sum_{m=1}^\infty\frac{1}{m}\frac{1}{2\sin\frac{2\pi m}{k}}\tr V^m.
\end{align}
Plugging this into \eqref{eq:open-GC}, we get
\begin{align}
e^{F^{\rm open}_{\rm pert}(\widehat{V})}&=1+\frac{e^{\frac{2\mu}{k}}}{2\sin\frac{2\pi}{k}}\tr V
+\frac{e^{\frac{4\mu}{k}}}{8\sin^2\frac{2\pi}{k}}(\tr V)^2
+\frac{ie^{\frac{4\mu}{k}}}{4\sin\frac{4\pi}{k}}\tr V^2\\
&\hspace{0.5cm}+\frac{e^{\frac{6\mu}{k}}}{48\sin^3\frac{2\pi}{k}}(\tr V)^3
+\frac{ie^{\frac{6\mu}{k}}}{8\sin\frac{2\pi}{k}\sin\frac{4\pi}{k}}\tr V \tr V^2
-\frac{e^{\frac{6\mu}{k}}}{6\sin\frac{6\pi}{k}}\tr V^3+\cdots.\notag
\end{align}
Therefore we immediately find
\begin{align}
\vev{\Str U^n}^{\rm GC(pert)}
=\frac{i^{n-1}}{2\sin\frac{2\pi n}{k}}e^{\frac{2n\mu}{k}} ,
\end{align}
and the factorization property
\begin{align}
\vev{(\Str U)^{n_1}(\Str U^2)^{n_2}\cdots}^{\rm GC(pert)}
=(\vev{\Str U}^{\rm GC(pert)})^{n_1}
(\vev{\Str U^2}^{\rm GC(pert)})^{n_2}\cdots.
\label{eq:factorization}
\end{align}
Note that this factorization property does not hold if the instanton
effect is taken into account.
One can check that these results reproduce \eqref{hookpert} for the
hook representations.

From the factorization property \eqref{eq:factorization}, one finds
that the perturbative part of the half BPS Wilson loop in the
representation $\mathbf{R}$ scales as
\begin{align}
\vev{W_{\mathbf{R}}}^{\rm GC(pert)}\sim e^{\frac{2n\mu}{k}} ,
\end{align}
where $n$ is the number of boxes of Young diagram $\mathbf{R}$ and
we have dropped the prefactor independent of $\mu$.
Coming back to the VEV in the canonical ensemble via \eqref{eq:GC-C},
we find that the perturbative part of the half BPS Wilson loop in
arbitrary representation gives the following Airy function behavior
\begin{align}
\vev{W_{\mathbf{R}}}_N^{\rm (pert)}
\sim {\rm Ai} \Biggl[ \left( \frac{2}{\pi^2 k} \right)^{-1/3}
\left( N -\frac{k}{24} -\frac{6n+1}{3k} \right)  \Biggr] ,
\end{align}
where the proportional coefficient depends only on $k$.
From this expression, we can also find the large $N$ limit as
\begin{align}
\frac{\vev{W_{\mathbf{R}}}_N}{Z(N)}
\sim e^{n \pi \sqrt{2\lambda}} \qquad (N \to \infty),
\end{align}
where $\lambda=N/k$ is the 't Hooft coupling.
Note that this exponent is the same as $n$ times of an classical
string action on the gravity side \cite{DT}.

\subsection{Worldsheet instantons}
Let us consider the worldsheet instanton corrections.
We first denote the general open string amplitude by
\begin{align}
F^{\rm open}(t,V)=\sum_{h=1}^\infty \sum_{\boldsymbol{\ell}} \sum_{m=1}^\infty
{\cal A}_{\ell_1,\dots,\ell_h}^{(m)} \tr V^{m \ell_1}\cdots \tr V^{m \ell_h}.
\label{eq:open_top2}
\end{align}
with
\begin{align}
{\cal A}_{\ell_1,\dots,\ell_h}^{(m)}
&=\sum_{\beta}\sum_{g=0}^\infty\frac{1}{h!}
n_{g,\beta,\boldsymbol{\ell}}\frac{1}{m}
\(2\sinh \frac{m g_{\rm top}}{2}\)^{2g-2}
\prod_{j=1}^h\(\frac{2}{\ell_j}\sinh\frac{m\l_j g_{\rm top}}{2}\)
e^{-m \beta\cdot t}.
\end{align}
After specifying $\beta=(d_1,d_2)$ and take the ``diagonal'' sum for
the open GV invariants
\begin{align}
n_{g,d,\boldsymbol{\ell}}
=\sum_{d_1+d_2=d} n_{g,(d_1,d_2),\boldsymbol{\ell}},
\end{align}
this becomes
\begin{align}
{\cal A}_{\ell_1,\dots,\ell_h}^{(m)}
&=\sum_{d=0}^\infty\sum_{g=0}^\infty\frac{(-1)^{g-1}}{h!}
n_{g,d,\boldsymbol{\ell}}\frac{1}{m}\(2\sin\frac{2\pi m}{k}\)^{2g-2}
\prod_{j=1}^h\(\frac{2i}{\ell_j}\sin\frac{2\pi m\l_j}{k}\)Q^{md}.
\label{eq:calA}
\end{align}
Thus we find from \eqref{eq:open_top2}, for example,
\begin{align}
\vev{\Str U}^{\rm GC(pert+WS)}\tr V
&={\cal A}_1^{(1)}\tr\widehat{V},\nn
\frac{1}{2}\vev{\Str U^2}^{\rm GC(pert+WS)}\tr V^2
&=({\cal A}_2^{(1)}+{\cal A}_1^{(2)})\tr\widehat{V}^2,\nn
\frac{1}{2}\vev{(\Str U)^2}^{\rm GC(pert+WS)}(\tr V)^2
&=\({\cal A}_{1,1}^{(1)}+\frac{1}{2}({\cal A}_1^{(1)})^2\)(\tr\widehat{V})^2,
\end{align}
where the relation between $V$ and $\widehat{V}$ is given by
\eqref{eq:hV-V}.

Using the explicit values of the open GV invariants listed in Tables 1
and 2 of \cite{GKM}, we obtain the worldsheet instanton corrections up
to order $Q^5$,
\begin{align}
&\vev{\Str U}^{\rm GC(pert+WS)}
=\frac{e^{\frac{2\mu}{k}}}{2\sin\frac{2\pi}{k}}\biggl[
1+2Q+3Q^2+10Q^3+\left(49-32\sin^2\frac{2\pi}{k}\right)Q^4\nn
&\hspace{1cm}+\left(288-576\sin^2\frac{2\pi}{k}
+352\sin^4\frac{2\pi}{k}\right)Q^5+\cO(Q^6)
\biggr],\nn
&\vev{\Str U^2}^{\rm GC(pert+WS)}
=\frac{ie^{\frac{4\mu}{k}}}{\sin\frac{4\pi}{k}\sin^2\frac{2\pi}{k}}\biggl[
\frac{1}{2}\sin^2\frac{2\pi}{k}+\frac{1}{2}\sin^2\frac{4\pi}{k}Q\nn
&\hspace{1cm}+\left(\sin^2\frac{2\pi}{k}+\sin^2\frac{4\pi}{k}\right)Q^2
+4\sin^2\frac{4\pi}{k}Q^3 \nn
&\hspace{1cm}+\left(\frac{3}{2}\sin^2\frac{2\pi}{k}
+18\sin^2\frac{4\pi}{k}-14\sin^2\frac{2\pi}{k}\sin^2\frac{4\pi}{k}
\right)Q^4\nn
&\hspace{1cm}+\biggl(104-224\sin^2\frac{2\pi}{k}
+160\sin^4\frac{2\pi}{k}\biggr)\sin^2\frac{4\pi}{k}Q^5+\cO(Q^6)
\biggr] ,\nn
&\vev{(\Str U)^2}^{\rm GC(pert+WS)}
=\frac{e^{\frac{4\mu}{k}}}{\sin^2\frac{2\pi}{k}}\biggl[
\frac{1}{4}+\left(1-\sin^2\frac{2\pi}{k}\right)Q
+\left(\frac{5}{2}-2\sin^2\frac{2\pi}{k}\right)Q^2\nn
&\hspace{1cm}+\left(8-8\sin^2\frac{2\pi}{k}\right)Q^3
+\left(\frac{147}{4}-64\sin^2\frac{2\pi}{k}+28\sin^4\frac{2\pi}{k}\right)Q^4\nn
&\hspace{1cm}+\left(208-656\sin^2\frac{2\pi}{k}
+768\sin^4\frac{2\pi}{k}-320\sin^6\frac{2\pi}{k}\right)Q^5+\cO(Q^6)
\biggr].
\label{eq:WS_GKM}
\end{align}

As discussed in \cite{AKV}, the $g=0$ terms of
$\vev{\Str U}^{\rm GC(pert+WS)}$ are given by the factor
$(Q/z)^{\frac{1}{2}}$ representing the worldsheet instanton
corrections to the disk amplitude.
Here $z$ and $Q$ are related by the mirror map of local
$\mathbb{P}^1\times\mathbb{P}^1$ along the diagonal slice $z_1=z_2=z$
\begin{align}
\frac{1}{2}\log\frac{Q}{z}
=2z\,{}_4F_3\left(1,1,\frac{3}{2},\frac{3}{2};2,2,2;16z\right).
\end{align}
Inverting this relation, the worldsheet instanton corrections to
the disk amplitude are found to be
\begin{align}
f(Q) &\equiv  \left(\frac{Q}{z}\right)^{\frac{1}{2}}\nn
&=1+2Q+3Q^2+10Q^3+49Q^4+288Q^5+1892Q^6+13390Q^7+\cdots,
\end{align}
which reproduce the invariants $n_{g=0,d,(1)}$ listed in \cite{GKM}.
Interstingly, we find from \cite{MV} that the VEV of the Wilson loop
with widing $n$ is generically written as the following form,
\begin{align}
\vev{\Str U^n}^{\rm GC(pert+WS)}=e^{\frac{2n\mu}{k}}f(Q^n)\sum_{g,d} \sum_{n=\ell m}
N^g_{\vec{e}_m,d} \left( 2\sin \frac{2\pi \ell}{k}\right)^{2g-2}
\left( 2\sin \frac{2\pi n}{k} \right) Q^{d\ell}
\label{eq:winding-n}
\end{align}
where $N^g_{\vec{e}_m,d}$ are integers, which are related to the open
GV invariants $n_{g,d,\bm{\ell}}$.
Instead of the topological string consideration, we can also fix such
integers by comparing the matrix model results \cite{MP1,DMP1,KMSS} in
the 't Hooft limit because the genus expansion in this limit captures
all the worldsheet instanton corrections.
The similar comparison on the worldsheet instanton corrections to the
grand potential has been done in \cite{HMO2}.
In this way, we have fixed the values of $N^g_{\vec{e}_m,d}$ in the
very first few cases.
The result is summarized in Table~\ref{tab:list-N}.
For $n=1,2$, one can check that \eqref{eq:winding-n} with
Table~\ref{tab:list-N} indeed reproduce \eqref{eq:WS_GKM}.

\begin{table}[tb]
\caption{The values of $N^g_{\vec{e}_m,d}$.}
\begin{center}
\begin{tabular}{cc}

\begin{tabular}{rrccccc}
\hline
$N^g_{\vec{e}_1,d}$ & $d=0$ & $1$ & $2$ & $3$ & $4$ & $5$ \\
\hline
$g=0$ & $1$ & $0$ & $0$ & $0$ & $0$ & $0$  \\
$1$ & $0$ & $0$ & $0$ & $0$ & $-8$  & $-128$ \\
$2$ & $0$ & $0$ & $0$ & $0$ & $0$  & $22$ \\
\hline
\end{tabular}

&

\begin{tabular}{rrccc}
\hline
$N^g_{\vec{e}_2,d}$ & $d=0$ & $1$ & $2$ & $3$ \\
\hline
$g=0$ & $0$ & $1$ & $2$ & $6$  \\
$1$ & $0$ & $0$ & $0$ & $0$   \\
\hline
\end{tabular}

\vspace{0.5cm}
\\

\begin{tabular}{rrccc}
\hline
$N^g_{\vec{e}_3,d}$ & $d=0$ & $1$ & $2$ & $3$ \\
\hline
$g=0$ & $0$ & $1$ & $3$ & $9$  \\
$1$ & $0$ & $0$ & $0$ & $0$   \\
\hline
\end{tabular}

&

\begin{tabular}{rrccc}
\hline
$N^g_{\vec{e}_4,d}$ & $d=0$ & $1$ & $2$ & $3$ \\
\hline
$g=0$ & $0$ & $1$ & $4$ & $14$  \\
$1$ & $0$ & $0$ & $-4$ & $-8$   \\
\hline
\end{tabular}

\end{tabular}
\end{center}
\label{tab:list-N}
\end{table}

In the next section, we will confirm that these worldsheet instanton
corrections are indeed consistent with our numerical results.

\section{Numerical study and membrane instantons}
In this section, we numerically evaluate the VEVs of the half BPS
Wilson loops in hook representations by using the formulation
presented in sections \ref{sec:hook} and \ref{sec:nonhook}.
The main motivation of this analysis is to explore the membrane
instanton effects, which are very hard to be described in the
topological string theory.
The similar analysis has  been already done for the grand partition
function in \cite{HMO2,HMO3}.
We compute the VEVs in various hook representations for some values of
$k$.
Here we propose that the membrane instanton corrections are completely
encoded by the replacement $\mu \to \mu_{\rm eff}$ in the perturbative
part and the worldsheet instanton part as in \eqref{eq:Wnp}.
The effective chemical potential $\mu_{\rm eff}$ is explicitly given
\cite{HMO3} for even $k=2n$ as
\begin{equation}
\mu_{\rm eff}=\mu+(-1)^{n-1}2e^{-2\mu}\ _4 F_3
\left(  1,1,\frac{3}{2},\frac{3}{2};2,2,2; (-1)^n 16e^{-2\mu} \right) ,
\label{eq:mu-mu_eff-evenk}
\end{equation}
and conjectured for odd $k$ as
\begin{equation}
\mu_{\rm eff}=\mu+e^{-4\mu}\ _4 F_3
\left(  1,1,\frac{3}{2},\frac{3}{2};2,2,2;  -16e^{-4\mu} \right) .
\label{eq:mu-mu_eff-oddk}
\end{equation}
Below, we will check the proposal \eqref{eq:Wnp} by the numerical study.

\subsection{A procedure}
Let us consider the VEV for the half BPS Wilson loops in the
hook representation $(a|l)$.
The VEV is given by \eqref{hook},
\begin{align}
\vev{W_{(a|l)}}^{\rm GC}
&=\int\frac{dx dy}{(2\pi k)^2}
\bra a|x\ket\bra x|\frac{z}{1+z\rho_1}|y\ket\bra y|l\ket.
\end{align}
Let us first note that, the complex phase dependence only come from
$\bra a|x\ket$ and $\bra y|l\ket$, which is trivially given in
\eqref{ndef}, 
\begin{align}
\vev{W_{(a|l)}}^{\rm GC}
=e^{\frac{a(a+1)\pi i}{k}-\frac{l(l+1)\pi i}{k}}|\vev{W_{(a|l)}}^{\rm GC}|.
\end{align}
Hence we define a real function ${\mathcal W}_{(a|l)}$ with its
series expansion ${\mathcal W}_{(a|l)}^{(m)}$ as
\begin{align}
{\mathcal W}_{(a|l)}
\equiv|\vev{W_{(a|l)}}^{\rm GC}|
=\sum_{m=0}^\infty(-1)^mz^{m+1}{\cal W}_{(a|l)}^{(m)},
\label{eq:Wal_exp}
\end{align}
where ${\mathcal W}_{(a|l)}^{(m)}$ is given by
\begin{align}
{\cal W}_{(a|l)}^{(m)}
=\int\frac{dx dy}{(2\pi k)^2}\widehat{f}_a(x)
\rho_1^{m}(x,y)\widehat{f}_l(y),\quad
\widehat{f}_n(x)\equiv\frac{e^{(n+\frac{1}{2})\frac{x}{k}}}{\sqrt{2\cosh\frac{x}{2}}}.
\label{eq:I}
\end{align}
Of course, the VEVs for $(a|l)$ and $(l|a)$ should be complex
conjugate to each other, therefore we immediately find
\begin{align}
{\mathcal W}_{(a|l)}={\mathcal W}_{(l|a)}.
\end{align}
Our task is to evaluate the integral \eqref{eq:I}.
This can be done as follows.
Let us introduce the function by
\begin{align}
\phi_l^{(m)}(x)=\frac{1}{\sqrt{2\cosh\frac{x}{2}}}
\int\frac{dy}{2\pi k}\rho_1^m(x,y)\widehat{f}_l(y).
\end{align}
One easily finds that this function satisfies the recurrence relation
\begin{align}
\phi_l^{(m)}(x)=\frac{1}{2\cosh\frac{x}{2}}
\int\frac{dy}{2\pi k}\frac{1}{2\cosh\frac{x-y}{2k}}\phi_l^{(m-1)}(y),
\label{eq:int_eq}
\end{align}
with the initial condition
\begin{align}
\phi_l^{(0)}(x)=\frac{e^{(l+\frac{1}{2})\frac{x}{k}}}{2\cosh\frac{x}{2}}.
\label{eq:initial}
\end{align}
Once the function $\phi_l^{(m)}(x)$ is known, the integral
\eqref{eq:I} is easily evaluated as
\begin{align}
{\cal W}_{(a|l)}^{(m)}
=\int\frac{dx}{2\pi k}e^{(a+\frac{1}{2})\frac{x}{k}}\phi_l^{(m)}(x).
\label{eq:int_W}
\end{align}
We notice that the integral equation \eqref{eq:int_eq} is essentially
the same as that appearing in \cite{HMO1,PY,HMO2}.
One can solve it for any $k$ at least numerically.
Practically, we solve the integral equation up to a certain value $m=m_{\rm max}$,
and make an approximation for ${\mathcal W}_{(a|l)}$ \eqref{eq:Wal_exp} as
\begin{align}
{\mathcal W}_{(a|l)} \approx\sum_{m=0}^{m_{\rm max}}(-1)^mz^{m+1}{\cal W}_{(a|l)}^{(m)}.
\end{align}
Though this approximation is originally valid only for the small $\mu$ regime,
we extrapolate the profile to a reasonably large $\mu$ regime and 
fit the expansion coefficients of ${\mathcal W}_{(a|l)}$ by exact values at this regime.
This is the same strategy as that in \cite{HMO2}.

Before closing this subsection, we will briefly comment on the
convergence of integral.
In \eqref{eq:initial} and \eqref{eq:int_W}, there appear the
exponential factors that diverge in large $x$ limit.
Due to these factors, the integral \eqref{eq:int_W} converges only if
\begin{align}
k>2(a+l+1)=2|\mathbf{R}_{\rm hook}|,
\end{align}
where $|\mathbf{R}_{\rm hook}|$ is the size of Young diagram
corresponding to the hook representation $\mathbf{R}_{\rm hook}$.
Therefore, the grand canonical VEVs, and 
correspondingly the canonical VEVs $\vev{W_{\mathbf{R}}}_N$, are well-defined only for such values of $k$,
though it is much harder to see it directly in $\vev{W_{\mathbf{R}}}_N$ due to the complex phases in the original expression \eqref{eq:W-VEV}.
Such a behavior has also been found for the multiple winding Wilson loop in
\cite{KMSS}.

\subsection{Fundamental representation}
The simplest representation is the fundamental representation $(0|0)$,
\begin{align}
\vev{W_{(0|0)}}^{\rm GC}={\mathcal W}_{(0|0)},
\end{align}
We would like to evaluate ${\mathcal W}_{(0|0)}$ numerically for some
values of $k$.
By solving the integral equation, we have performed the numerical
computation, and find the non-perturbative corrections to
${\mathcal W}_{(0|0)}$  for $k=3,4,6,8,12$.
The results are as follows:
\begin{align}
{\mathcal W}_{(0|0)}|_{k=3}
&={\mathcal W}_{(0|0)}^{\rm (pert)}
\left(1+2Q+3Q^2+\frac{28}{3}Q^3+\frac{79}{3}Q^4+60Q^5
+\frac{1562}{9}Q^6+\cO(Q^7)\right), \nn
{\mathcal W}_{(0|0)}|_{k=4}
&={\mathcal W}_{(0|0)}^{\rm (pert)}
\left(1+2Q+2Q^2+12Q^3+22Q^4+124Q^5+276Q^6+\cO(Q^7)\right),\nn
{\mathcal W}_{(0|0)}|_{k=6}
&={\mathcal W}_{(0|0)}^{\rm (pert)}
\left(1+2Q+3Q^2+\frac{28}{3}Q^3+\frac{79}{3}Q^4+60Q^5
+\frac{1562}{9}Q^6+\cO(Q^7)\right), \nn
{\mathcal W}_{(0|0)}|_{k=8}
&={\mathcal W}_{(0|0)}^{\rm (pert)}
\left(1+2Q+3Q^2+10Q^3+\frac{65}{2}Q^4+89Q^5
+\frac{465}{2}Q^6+\cO(Q^7)\right), \nn
{\mathcal W}_{(0|0)}|_{k=12}
&={\mathcal W}_{(0|0)}^{\rm (pert)}
\left(1+2Q+3Q^2+10Q^3+41Q^4+166Q^5
+\frac{1844}{3}Q^6+\cO(Q^7)\right),
\label{eq:num_res1}
\end{align}
where ${\mathcal W}_{(0|0)}^{\rm (pert)}$ is 
\begin{align}
{\mathcal W}_{(0|0)}^{\rm (pert)}=\frac{e^{\frac{2\mu}{k}}}{2\sin\frac{2\pi}{k}}.
\label{eq:W_00}
\end{align}
To obtain the numerical results, we have chosen $m_{\rm max}$ to be the best value of the numerical fitting 
at each instanton number.
This best value decreases as the instanton number increases because of the exponential suppression
of the corrections and the numerical errors.
Anyway, for all the above cases, we have chosen $m_{\rm max} \sim 10$,
and the
numerical results of the coefficients match to these exact values \eqref{eq:num_res1}
in about $5$-digit accuracy.  
Especially, if we make a wrong guess at a certain instanton order, the next
order of instanton would grow exponentially, which could make our fitting totally
impossible.
Note that for the above values of $k$ the coefficients of \eqref{eq:W_00} become
particularly simple (especially rational) because the root of unity $e^{2\pi i/k}$ takes
the simple values. This is why we choose the above
values of $k$ for our fitting problem.

Let us compare these results with the theoretical prediction.
The worldsheet instanton corrections of
$\vev{W_{(0|0)}}^{\rm GC}=\vev{\Str U}^{\rm GC}$ are given by
\eqref{eq:WS_GKM}.
As mentioned before, we propose that the membrane instanton correction
can be incorporated by the replacement $\mu\to\mu_{\rm eff}$ in the
worldsheet instanton correction.
Thus our conjecture, including the membrane instanton effects, is
\begin{align}
{\mathcal W}_{(0|0)}
&=\frac{e^{\frac{2\mu_{\rm eff}}{k}}}{2\sin\frac{2\pi}{k}}\biggl[
1+2Q_{\rm eff}+3Q_{\rm eff}^2
+10Q_{\rm eff}^3+\left( 49-32\sin^2\frac{2\pi}{k}\right)Q_{\rm eff}^4\nn
&\hspace{1cm}+\left(288-576\sin^2\frac{2\pi}{k}
+352\sin^4\frac{2\pi}{k}\right)Q_{\rm eff}^5+\cO(Q_{\rm eff}^6)\biggr],
\label{eq:prediction}
\end{align}
where $Q_{\rm eff}=-e^{-\frac{4\mu_{\rm eff}}{k}}$.
For the comparison, we need to rewrite it in terms of
$Q=-e^{-\frac{4\mu}{k}}$.
Using the relations \eqref{eq:mu-mu_eff-evenk} and
\eqref{eq:mu-mu_eff-oddk} between $\mu$ and $\mu_{\rm eff}$, we find
\begin{align}
Q_{\rm eff}=\begin{cases} Q+\frac{4}{3}Q^4+\cO(Q^7) \quad &(k=3,6) \\
Q+2Q^3+11Q^5+\cO(Q^7) \quad&(k=4) \\
Q+Q^5+\cO(Q^9) \quad&(k=8) \\
Q+\frac{2}{3}Q^7+\cO(Q^{13}) \quad&(k=12).
\end{cases}
\label{eq:Q-Qeff}
\end{align}
Plugging these into \eqref{eq:prediction}, one can check that the
corrections exactly agree with the numerical ones \eqref{eq:num_res1}
up to $Q^5$.
We emphasize that only the worldsheet instanton correction does not
explain the numerical results \eqref{eq:num_res1}.
We need to replace $\mu$ by $\mu_{\rm eff}$ to reproduce them.
This is due to the membrane instanton effects.

\subsection{Young diagrams with two boxes}
There are two representations with two-box Young diagrams:
\begin{align}
\vev{W_{(1|0)}}^{\rm GC}=e^{\frac{2\pi i}{k}}{\mathcal W}_{(1|0)},\quad 
\vev{W_{(0|1)}}^{\rm GC}=e^{-\frac{2\pi i}{k}}{\mathcal W}_{(0|1)}.
\end{align}
We have the relation ${\mathcal W}_{(1|0)}={\mathcal W}_{(0|1)}$.
By the similar computation to the fundamental representation, we find
\begin{align}
{\mathcal W}_{(1|0)}|_{k=6}&={\mathcal W}_{(1|0)}^{\rm (pert)}
\left(1+Q+4Q^2+\frac{20}{3}Q^3+18Q^4+\frac{172}{3}Q^5
+\frac{1190}{9}Q^6+\cO(Q^7)\right),\nn
{\mathcal W}_{(1|0)}|_{k=8}&={\mathcal W}_{(1|0)}^{\rm (pert)}
\left(1+2Q+6Q^2+16Q^3+46Q^4+128Q^5+364Q^6+\cO(Q^7)\right),\nn
{\mathcal W}_{(1|0)}|_{k=12}&={\mathcal W}_{(1|0)}^{\rm (pert)}
\left(1+3Q+8Q^2+24Q^3+90Q^4+348Q^5+\frac{3862}{3}Q^6+\cO(Q^7)\right),
\label{eq:num_res2}
\end{align}
with 
\begin{align}
{\mathcal W}_{(1|0)}^{\rm (pert)}
=\frac{e^{\frac{4\mu}{k}}}{4\sin\frac{2\pi}{k}\sin\frac{4\pi}{k}}.
\end{align}

Let us also compare these results with our prediction.
Note that $\vev{W_{(1|0)}}^{\rm GC}$ is given by
\begin{align}
\vev{W_{(1|0)}}^{\rm GC}
=\frac{1}{2}\vev{(\Str U)^2}^{\rm GC}+\frac{1}{2}\vev{\Str U^2}^{\rm GC} .
\end{align}
The worldsheet instanton corrections of $\vev{(\Str U)^2}^{\rm GC}$
and $\vev{\Str U^2}^{\rm GC}$ are given by \eqref{eq:WS_GKM}.
Thus our prediction is
\begin{align}
{\mathcal W}_{(1|0)}
&=\frac{e^{\frac{4\mu_{\rm eff}}{k}}}{4\sin^3\frac{2\pi}{k}\sin\frac{4\pi}{k}}\biggl[
\sin^2\frac{2\pi}{k}+\sin^2\frac{4\pi}{k}Q_{\rm eff}
+\(2\sin^2\frac{2\pi}{k}+2\sin^2\frac{4\pi}{k}\)Q_{\rm eff}^2\nn
&+8\sin^2 \frac{4\pi}{k}Q_{\rm eff}^3
+\(3\sin^2\frac{2\pi}{k}+36\sin^2\frac{4\pi}{k}
-28\sin^2\frac{2\pi}{k}\sin^2\frac{4\pi}{k}\)Q_{\rm eff}^4\nn
&+\(208-448\sin^2\frac{2\pi}{k}
+320\sin^4\frac{2\pi}{k}\)\sin^2\frac{4\pi}{k}Q_{\rm eff}^5
+\cO(Q_{\rm eff}^6)\biggr].
\label{eq:prediction2}
\end{align}
Using the relation \eqref{eq:Q-Qeff},
one finds that the corrections again agree with the numerical ones
\eqref{eq:num_res2} 
up to $Q^5$.

\subsection{Young diagrams with three boxes}
For the three-box Young diagrams, there are two non-trivial real
functions,
\begin{align}
\vev{W_{(2|0)}}^{\rm GC}
&=e^{\frac{6\pi i}{k}}{\mathcal W}_{(2|0)},\quad
\vev{W_{(1|1)}}^{\rm GC}={\mathcal W}_{(1|1)},\quad
\vev{W_{(0|2)}}^{\rm GC}=e^{-\frac{6\pi i}{k}}{\mathcal W}_{(0|2)},
\end{align}
with the constraint ${\mathcal W}_{(2|0)}={\mathcal W}_{(0|2)}$.
From the numerical analysis, we find
\begin{align}
{\mathcal W}_{(2|0)}|_{k=8}
&={\mathcal W}_{(2|0)}^{\rm (pert)}
\(1+Q^2+8Q^3+\frac{33}{2}Q^4+40Q^5+\frac{235}{2}Q^6+\cO(Q^7)\),\nn
{\mathcal W}_{(2|0)}|_{k=12}
&={\mathcal W}_{(2|0)}^{\rm (pert)}
\(1+2Q+8Q^2+32Q^3+116Q^4+426Q^5+1534Q^6+\cO(Q^7)\),
\end{align}
and
\begin{align}
{\mathcal W}_{(1|1)}|_{k=8}
&={\mathcal W}_{(1|1)}^{\rm (pert)}
\(1+2Q+5Q^2+14Q^3+\frac{73}{2}Q^4+105Q^5+\frac{591}{2}Q^6+\cO(Q^7)\),\nn
{\mathcal W}_{(1|1)}|_{k=12}
&={\mathcal W}_{(1|1)}^{\rm (pert)}\(1+4Q+12Q^2+38Q^3+136Q^4+508Q^5+1866Q^6+\cO(Q^7)\) .
\label{eq:W11_num}
\end{align}
Note that to compare these results with the theoretical prediction, we
need to know the open GV invariants $n_{g,d,\boldsymbol{\ell}}$ for
$\boldsymbol{\ell}=(3), (2,1), (1,1,1)$, whose explicit values are not
found in the literature.
Instead, one can compare the result for the Wilson loop with winding
$3$.
The VEV of the Wilson loop with winding $3$ is computed as
\begin{align}
\vev{W_3}^{\rm GC}&=\vev{W_{(2|0)}}^{\rm GC}-\vev{W_{(1|1)}}^{\rm GC}+\vev{W_{(0|2)}}^{\rm GC} \nn
&=2\cos \( \frac{6\pi}{k} \) {\mathcal W}_{(2|0)}-{\mathcal W}_{(1|1)},
\end{align}
From \eqref{eq:winding-n} with Table~\ref{tab:list-N}, on the other
hand, we obtain
\begin{align}
\vev{W_3}^{\rm GC}&=-e^{\frac{6\mu_{\rm eff}}{k}}\biggl[
\frac{1}{2} \csc\frac{6\pi}{k}
+\frac{1}{2} \csc^2 \frac{2\pi}{k} \sin \frac{6\pi}{k} Q_{\rm eff}
+\frac{3}{2} \csc^2 \frac{2\pi}{k} \sin \frac{6\pi}{k} Q_{\rm eff}^2 \nn
&\quad+\( \csc \frac{6\pi}{k}
+\frac{9}{2} \csc^2 \frac{2\pi}{k} \sin \frac{6\pi}{k}\)Q_{\rm eff}^3
+\cO(Q_{\rm eff}^4) \biggr].
\end{align}
One can check that this reproduces the above results for $k=8,12$ up to $Q^3$.

\subsection{Young diagrams with four boxes}
For the four-box case, there are four hook representations and one
non-hook representation.
For the hook representations, we have
\begin{alignat}{2}
\vev{W_{(3|0)}}^{\rm GC}&=e^{\frac{12\pi i}{k}} {\mathcal W}_{(3|0)},
&\quad\vev{W_{(2|1)}}^{\rm GC}&=e^{\frac{4\pi i}{k}} {\mathcal W}_{(2|1)} ,\nn
\vev{W_{(0|3)}}^{\rm GC}&=e^{-\frac{12\pi i}{k}} {\mathcal W}_{(0|3)},
&\quad\vev{W_{(1|2)}}^{\rm GC}&= e^{-\frac{4\pi i}{k}}{\mathcal W}_{(1|2)}, 
\end{alignat}
with ${\mathcal W}_{(3|0)}={\mathcal W}_{(0|3)}$ and
${\mathcal W}_{(2|1)}={\mathcal W}_{(1|2)}$.
For the non-hook representation $(1,0|1,0)$, the VEV is given by the
determinant formula
\begin{align}
\vev{W_{(1,0|1,0)}}^{\rm GC}=\det\begin{pmatrix}
\vev{W_{(1|1)}}^{\rm GC}&\vev{W_{(1|0)}}^{\rm GC}\\ 
\vev{W_{(0|1)}}^{\rm GC}&\vev{W_{(0|0)}}^{\rm GC}
\end{pmatrix}
={\mathcal W}_{(1|1)}{\mathcal W}_{(0|0)}-{\mathcal W}_{(1|0)}{\mathcal W}_{(0|1)}.
\end{align}
From the numerical analysis, we find
\begin{align}
{\mathcal W}_{(3|0)}|_{k=12}
&={\mathcal W}_{(3|0)}^{\rm (pert)}
\(1+Q^2+12Q^3+61Q^4+216Q^5+\frac{1417}{2}Q^6+\cO(Q^7)\),\nn
{\mathcal W}_{(2|1)}|_{k=12}
&={\mathcal W}_{(2|1)}^{\rm (pert)}
\(1+3Q+10Q^2+36Q^3+133Q^4+486Q^5+\frac{5258}{3}Q^6+\cO(Q^7)\),
\end{align}
One can check that the VEV of the Wilson loop with winding $4$ at
$k=12$ is reproduced from these results.

\subsection{Implications}
The grand canonical VEVs of the half BPS Wilson loops are in general
complex.
As was seen before, however, their phase dependences are trivial.
This fact implies that there are some non-trivial relation among open
GV invariants $n_{g,d,\boldsymbol{\ell}}$ for different $\boldsymbol{\ell}$.
Let us see this here.
In the size $2$ representations, we have
\begin{align}
\vev{(\Str U)^2}^{\rm GC}
&=\vev{W_{(1|0)}}^{\rm GC}+\vev{W_{(0|1)}}^{\rm GC}
=2\cos\frac{2\pi}{k}{\mathcal W}_{(1|0)},\nn
\vev{\Str U^2}^{\rm GC}
&=\vev{W_{(1|0)}}^{\rm GC}-\vev{W_{(0|1)}}^{\rm GC}
=2i\sin\frac{2\pi}{k}{\mathcal W}_{(1|0)},
\end{align}
where we have used ${\mathcal W}_{(1|0)}={\mathcal W}_{(0|1)}$.
These expressions immediately lead to the exact relation,
\begin{align}
\frac{\vev{\Str U^2}^{\rm GC}}{\vev{(\Str U)^2}^{\rm GC}}
=i\tan\frac{2\pi}{k}.
\label{eq:rel-2dim}
\end{align}
This relation gives a non-trivial relation among the open GV
invariants $n_{g,d,(1)}$, $n_{g,d,(2)}$ and $n_{g,d,(1,1)}$.
For very lower $g$ and $d$, we find
\begin{align}
&n_{0,1,(2)}=n_{0,1,(1,1)}=\frac{n_{0,1,(1)}}{2},\nn
&n_{0,2,(2)}=n_{0,2,(1,1)}=\frac{1}{4}(2n_{0,2,(1)}+n_{0,1,(1)}^2-n_{0,1,(1)}),\nn
&n_{0,3,(2)}=n_{0,3,(1,1)}=\frac{1}{2}(n_{0,3,(1)}+n_{0,2,(1)}n_{0,1,(1)}),\nn
&n_{0,4,(2)}=\frac{1}{4}(2n_{0,4,(1)}+2n_{0,3,(1)}n_{0,1,(1)}+n_{0,2,(1)}^2-n_{0,2,(1)}), \nn
&n_{0,4,(1,1)}-4 n_{1,4,(1,1)}=\frac{1}{4}(2n_{0,4,(1)}+2n_{0,3,(1)}n_{0,1,(1)}+n_{0,2,(1)}^2-n_{0,2,(1)}-8n_{1,4,(1)}),\nn
&n_{1,4,(1,1)}=n_{1,4,(2)}.
\end{align}
One can check that the expressions \eqref{eq:WS_GKM} indeed satisfy
the relation \eqref{eq:rel-2dim} 
up to order $Q^5$.

Similarly, from the relation
\begin{align}
\frac{1}{3}\vev{(\Str U)^3}^{\rm GC}+\frac{2}{3}\vev{\Str U^3}^{\rm GC}
&=2\cos\frac{6\pi}{k}{\mathcal W}_{(2|0)},\nn
\vev{\Str U\Str U^2}^{\rm GC}&=2i\sin\frac{6\pi}{k}{\mathcal W}_{(2|0)},
\end{align}
we find
\begin{align}
\vev{\Str U\Str U^2}^{\rm GC}
=i\tan \frac{6\pi}{k}\left(
\frac{1}{3}\vev{(\Str U)^3}^{\rm GC}+\frac{2}{3}\vev{\Str U^3}^{\rm GC}
\right).
\end{align}
This gives an non-trivial relation among $n_{g,d,\boldsymbol{\ell}}$
for $\boldsymbol{\ell}=(1), (2), (1,1), (3), (2,1), (1,1,1)$.

Also, the Giambelli formula \eqref{nonhook} gives non-trivial
relations among the open GV invariants.
For the representation $\mathbf{R}=(1,0|1,0)$, we find the relation
\begin{align}
&\frac{1}{12} \vev{(\Str U)^4}^{\rm GC}
-\frac{1}{3}\vev{\Str U \Str U^3}^{\rm GC}
+\frac{1}{4}\vev{(\Str U^2)^2}^{\rm GC} \nonumber \\
&=\frac{1}{3} \vev{(\Str U)^3}^{\rm GC}\vev{\Str U}^{\rm GC}
-\frac{1}{4}(\vev{(\Str U)^2}^{\rm GC})^2 \nonumber\\
& \ \  -\frac{1}{3}\vev{\Str U}^{\rm GC}\vev{\Str U^3}^{\rm GC}
  +\frac{1}{4}(\vev{\Str U^2}^{\rm GC})^2, 
\end{align}
or equivalently, in terms of ${\cal A}_{\ell_1,\dots,\ell_h}^{(m)}$
defined by \eqref{eq:calA}, the relation is written as
\begin{align}
{\cal A}_{1,1,1,1}^{(1)}-{\cal A}_{3,1}^{(1)}+{\cal A}_{2,2}^{(1)}
=-{\cal A}_{1,1}^{(2)}-({\cal A}_{1,1}^{(1)})^2.
\end{align}
Substituting \eqref{eq:calA} into this, we obtain the relation among
the open GV invariants.\footnote{
To capture the membrane instanton correction, we need to replace $Q$
in \eqref{eq:calA} by $Q_{\rm eff}$, but this replacement does not
change the relations at all.}

\section{Conclusion}
In this paper we have proposed the Fermi gas formalism for the VEVs of
the half BPS Wilson loops in arbitrary representations.
For the case of the hook representations, we present the formula in
terms of the convolution of integrations.
For the case of the non-hook representations, we reduce the
computation to the hook case by a determinant formula similar to the
Giambelli formula for the Schur polynomial.
After working out these expressions for the VEVs, we also present a
numerical study.
We find that besides the worldsheet instanton corrections we also have
the membrane instanton corrections which can be incorporated by
shifting the chemical potential $\mu$ into $\mu_{\rm eff}$ as we did
in studying the bound states in the ABJM partition function.

We conclude our paper by listing several discussions on the further
directions.

Based on the numerical results, we conclude that the membrane
instanton correction is completely encoded in the perturbative and the
worldsheet instanton parts by replacing $\mu$ by $\mu_{\rm eff}$.
Let us recall that in the partition function, there is also a pure
membrane instanton correction, as well as the bound states of the
worldsheet instantons and the membrane instantons.
This pure membrane instanton correction is directly related to the
non-perturbative effect in the closed topological string \cite{HMMO}
(see also \cite{Lockhart:2012vp}).
Our Wilson loop result \eqref{eq:Wnp} implies that there seem to be no
pure membrane instanton corrections in the open topological string on
``diagonal'' local $\mathbb{P}^1 \times \mathbb{P}^1$.
It would be interesting to confirm this in the topological string
framework.

Most of our analysis here focus on the half BPS Wilson loops, which
have nice counterparts in the topological string.
Our method presented in section 2, however, can be applicable to the
$1/6$ BPS Wilson loops.
The topological string counterparts to such $1/6$ BPS Wilson loops are
unclear, thus it would be important to reveal the structure of
instanton effects in the $1/6$ BPS Wilson loops by using our method.
It is also interesting to perform Monte Carlo simulation \cite{W_KEK}
of the $1/6$ BPS Wilson loops in low dimensional representations,
which has been useful for the partition function \cite{KEK}.
It would also be illuminating to apply our formalism to other
observables in the ABJM theory such as the vortex loop
\cite{Drukker:2012sr} and energy-momentum tensor correlator
\cite{Closset:2012ru}, which can be also simplified by the
localization method.

In the topological string theory, we have a set of open GV invariants
for each representation.
In our Fermi gas formalism, we find several non-trivial relations
among such invariants.
The simplest one is the symmetry of taking the transpose in the Young
diagram.
For example, disregarding a difference in the trivial phase factor,
the VEVs of the half BPS ABJM Wilson loops in the symmetric and
anti-symmetric representations are equal with each other.
Hence this triviality of the phase factor imposes highly non-trivial
relations in the open GV invariants.
The origin of this property is unclear on the topological string side
at present.
Besides, the VEVs in the non-hook representations enjoy the Giambelli
property.
Technically, the Giambelli property imposes many interesting relations
and reduces largely the unknown open GV invariants.
Using the Giambelli property, we can show that the number of unknown
GV invariants at each order reduces to the number of boxes $n$, which
originally increases with the number of representations, namely,
partitions
$p(n)\sim e^{\pi\sqrt{\frac{2n}{3}}}/(4\sqrt{3}n)$.
It is interesting to clarify what kind of relations the transposition
symmetry and the Giambelli compatibility will impose on the open GV
invariants.
We also ask whether these kinds of relations appear in more general
topological string theories or not.
Since we have studied only the local
$\mathbb{P}^1 \times \mathbb{P}^1$ topological string, the relations
might be accidental properties in this model.
If these are common in a class of topological strings, we expect that
there are some extra structures, which naturally explain the
relations.
For example, since the topological recursion of Eynard and Orantin
\cite{Eynard:2007kz} gives relations among all open string invariants,
this might explain the relations coming from the transposition
symmetry and the Giambelli compatibility.

A natural open question is the physical interpretation of the
Giambelli compatibility.
It would be nice to understand its meaning from the brane
configuration or the gravity analysis.
We hope that this would be a clue to understand M-theory.

\vskip5mm
\centerline{\bf Acknowledgements}
\vskip3mm
\noindent
We are grateful to Heng-Yu Chen, Nadav Drukker, Marcos Marino, Tomoki
Nosaka, Kazutoshi Ohta, Soo-Jong Rey, Masaki Shigemori, Takao Suyama
for useful discussions.
Especially we would like to thank Sho Matsomoto for sharing his idea
of the proof of Giambelli compatibity with us.
The work of Y.H. is supported in part by the JSPS Research Fellowship
for Young Scientists, while the work of K.O. is supported in part by
JSPS Grant-in-Aid for Young Scientists (B) \#23740178.

\end{document}